\documentclass[nofootinbib,superscriptaddress,notitlepage,amsmath,amssymb,fleqn,prd,preprintnumbers,nobalancelastpage,floatfix]{revtex4-1}
\usepackage{graphicx,tabularx,xspace,todonotes,mathtools,color}

\newcommand{\Chili}{{\sc Chili}\xspace}
\newcommand{\Comix}{{\sc Comix}\xspace}
\newcommand{\Sherpa}{{\sc Sherpa}\xspace}
\newcommand{\MadNIS}{{\sc MadNIS}\xspace}
\usepackage{hyperref}
\hypersetup{hidelinks,
  pdfauthor={Enrico Bothmann,Taylor Childers,Walter Giele,Florian Herren,Stefan Hoeche,Joshua Isaacsson,Max Knobbe,Rui Wang},
  pdftitle={Efficient phase-space generation for hadron collider event simulation},
  pdfkeywords={Phase-space integration}
}

\newcommand{\affANL}{Argonne National Laboratory, Lemont, IL, 60439, USA}
\newcommand{\affFermilab}{Fermi National Accelerator Laboratory, Batavia, IL 60510, USA}
\newcommand{\affGoettingenU}{Institut f\"ur Theoretische Physik, Georg-August-Universit\"at G\"ottingen, 37077 G\"ottingen, Germany}

\begin{document}
\preprint{FERMILAB-PUB-23-032-T, MCNET-23-02}
\title{Efficient phase-space generation for hadron collider event simulation}
\author{Enrico~Bothmann}\affiliation{\affGoettingenU}
\author{Taylor~Childers}\affiliation{\affANL}
\author{Walter~Giele}\affiliation{\affFermilab}
\author{Florian~Herren}\affiliation{\affFermilab}
\author{Stefan~H{\"o}che}\affiliation{\affFermilab}
\author{Joshua~Isaacson}\affiliation{\affFermilab}
\author{Max~Knobbe}\affiliation{\affGoettingenU}
\author{Rui~Wang}\affiliation{\affANL}

\begin{abstract}
  We present a simple yet efficient algorithm for phase-space integration at hadron colliders.
  Individual mappings consist of a single t-channel combined with any number of s-channel decays, 
  and are constructed using diagrammatic information. The factorial growth in the number
  of channels is tamed by providing an option to limit the number of s-channel topologies.
  We provide a publicly available, parallelized code in C++ and test its performance 
  in typical LHC scenarios.
\end{abstract}

\maketitle

\section{Introduction}
The problem of phase-space integration is omnipresent in particle physics.
Efficient methods to evaluate phase-space integrals are needed in order to predict
cross sections and decay rates for a variety of experiments, and they are required
for both theoretical calculations and event simulation.
In many cases, the integrand to be evaluated features a number of narrow peaks, 
corresponding to the resonant production of unstable massive particles.
In other cases, the integrand has intricate discontinuities, arising from 
cuts to avoid the singular regions of scattering matrix elements in theories
with massless force carriers, such as QED and QCD. In most interesting scenarios,
the phase space is high dimensional, such that analytic integration is ruled out,
and Monte-Carlo (MC) integration becomes the only viable option.

Many techniques have been devised to deal with this problem~\cite{James:1968gu,
  Byckling:1969luw,Byckling:1969sx,Kleiss:1985gy,Kanaki:2000ey,Maltoni:2002qb,
  vanHameren:2002tc,Gleisberg:2008fv,vanHameren:2010gg,Platzer:2013esa}. 
Among the most successful ones are factorization based approaches~\cite{
 James:1968gu,Byckling:1969luw,Byckling:1969sx}
and multi-channel integration techniques~\cite{Kleiss:1994qy}.
They allow to map the structure of the integral to the diagrammatic structure of
the integrand. For scalar theories, and ignoring the effect of phase-space cuts,
this corresponds to an ideal variable transformation. Realistic multi-particle
production processes are much more complex, both because of the non-scalar nature
of most of the elementary particles, and because of phase-space restrictions.
Adaptive Monte-Carlo methods~\cite{Lepage:1977sw,Ohl:1998jn,Lepage:2020tgj,
  Jadach:2002kn,Hahn:2004fe,vanHameren:2007pt} are therefore used by most theoretical
calculations and event generators to map out structures of the integrand which are
difficult to predict. More recently, neural networks have emerged as a promising 
tool for this particular task~\cite{Klimek:2018mza,Bothmann:2020ywa,Gao:2020vdv,
 Gao:2020zvv,Heimel:2022wyj,Maitre:2022xle,Verheyen:2022tov,Butter:2022rso}.

In this letter, we introduce a novel phase-space integrator which combines several
desirable features of different existing approaches while still remaining relatively simple.
In particular, we address the computational challenges discussed in a number of reports
of the HEP Software Foundation~\cite{
  HSFPhysicsEventGeneratorWG:2020gxw,Valassi:2020ueh,HSFPhysicsEventGeneratorWG:2021xti}
and the recent Snowmass community study~\cite{Campbell:2022qmc}, which emphasize
the importance of portable computing models.
Our algorithm is based on the highly successful integration techniques employed in 
MCFM~\cite{Campbell:2003hd,Campbell:1999ah,Campbell:2011bn,Campbell:2019dru}, combined with a
standard recursive approach for s-channel topologies as used in many modern
simulation programs. We provide a stand-alone implementation,
which we call \Chili (Common High-energy Integration LIbrary)\footnote{%
  The source code can be found at \url{https://gitlab.com/spice-mc/chili}.},
which includes the Vegas algorithm~\cite{Lepage:1977sw} and MPI parallelization.
We also implement Python bindings via nanobind~\cite{nanobind} 
and to Tensorflow~\cite{Abadi_TensorFlow_Large-scale_machine_2015}, 
providing an interface the normalizing-flow based neural network 
integration frameworks iFlow~\cite{Gao:2020vdv} and \MadNIS~\cite{Heimel:2022wyj}. 
To assess the performance of our new code, we combine it with the matrix-element
generators in the general-purpose event generator \Sherpa~\cite{Gleisberg:2008fv,Sherpa:2019gpd}
and devise a proof of concept for the computation of real-emission next-to-leading 
order corrections by adding a forward branching generator which makes use 
of the phase-space mappings of the Catani-Seymour dipole subtraction 
formalism~\cite{Catani:1996vz,Catani:2002hc}.

The outline of the paper is as follows: Section~\ref{sec:algorithms} discusses
the algorithms used in our new generator. Section~\ref{sec:performance} presents
performance measures obtained in combination with \Comix~\cite{Gleisberg:2008fv},
and Amegic~\cite{Krauss:2001iv}, and Sec.~\ref{sec:outlook} includes a summary and outlook.

\section{The Algorithm}
\label{sec:algorithms}
\begin{figure}
  \includegraphics[width=0.75\textwidth]{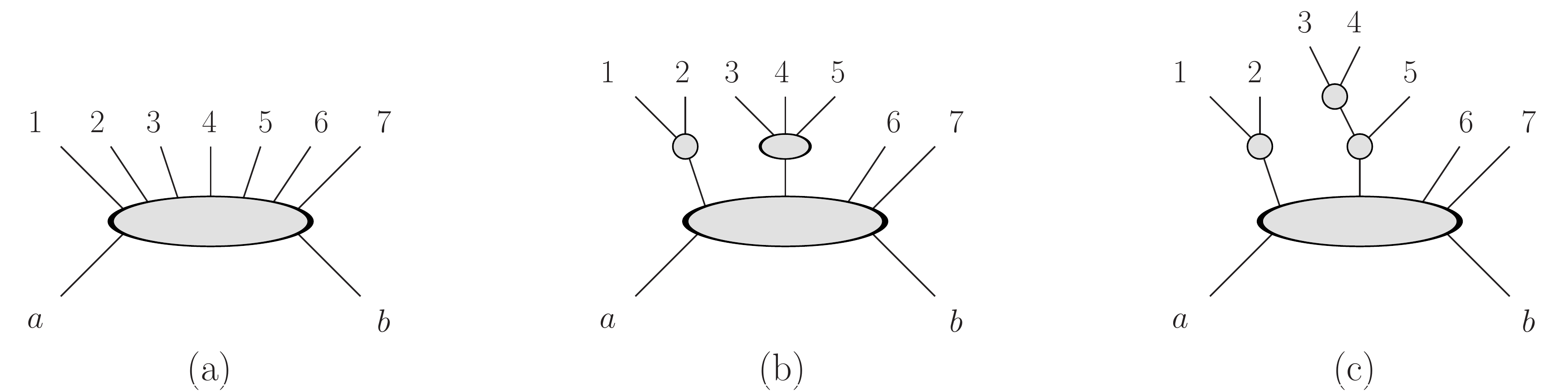}
  \caption{Example application of the phase-space factorization formula, Eq.~\eqref{eq:split_ps}. 
    Particles 1 through 7 are produced in the collision of particles $a$ and $b$.
    Figure~(a) represents a pure t-channel configuration, cf.\ Sec.~\ref{sec:building_blocks}.
    In Fig.~(b), the differential 7-particle phase-space element is factorized
    into the production of four particles, two of which are the pseudo-particles
    $\{1,2\}$ and $\{3,4,5\}$, which subsequently decay. In Fig.~(c),
    the decay of $\{3,4,5\}$ is again factorized into two consecutive decays.}
  \label{fig:factorization_example}
\end{figure}

One of the most versatile approaches to phase-space integration for high-energy
collider experiments is to employ the factorization properties of the $n$-particle 
phase-space integral~\cite{Byckling:1969sx}. Consider a $2\to n$ scattering process,
where we label the incoming particles by $a$ and $b$ and outgoing particles by $1\ldots n$.
The corresponding $n$-particle differential phase-space element reads
\begin{equation}\label{eq:n_particle_ps}
  {\rm d}\Phi_n(a,b;1,\ldots,n)=
    \left[\,\prod\limits_{i=1}^n\frac{{\rm d}^3\vec{p}_i}{(2\pi)^3\,2E_i}\,\right]\,
    (2\pi)^4\delta^{(4)}\bigg(p_a+p_b-\sum_{i=1}^n p_i\bigg)\;.
\end{equation}
Following Ref.~\cite{James:1968gu}, the full differential phase-space element can
be reduced to lower-multiplicity differential phase-space elements as follows:
\begin{equation}\label{eq:split_ps}
  {\rm d}\Phi_n(a,b;1,\ldots,n)=
    {\rm d}\Phi_{n-m+1}(a,b;\pi,m+1,\ldots,n)\,\frac{{\rm d} s_\pi}{2\pi}\,
    {\rm d}\Phi_m(\pi;1,\ldots,m)\;,
\end{equation}
where $\pi$ indicates an intermediate pseudo-particle of virtuality $s_\pi=p_\pi^2$.
Equation~\eqref{eq:split_ps} allows to compose the full differential phase-space
element from building blocks which correspond to a single t-channel 
production process and a number of s-channel decays, as depicted in
Fig.~\ref{fig:factorization_example}. By repeated application of
Eq.~\eqref{eq:split_ps}, all decays can be reduced to two-particle decays, 
with differential phase-space elements ${\rm d}\Phi_{2}$. This allows to match 
the structure of the phase-space integral onto the structure of the Feynman 
diagrams in the integrand at hand, a technique that is known as diagram-based
integration.

\subsection{The t- and s-channel building blocks}
\label{sec:building_blocks}
In this subsection, we first describe the techniques to perform the integration using
a pure t-channel differential phase-space element, ${\rm d}\Phi_n(a,b;1,\ldots,n)$.
The final-state momenta $p_1$ through $p_n$ can be associated with on-shell particles,
or they can correspond to intermediate pseudo-particles whose virtuality is an additional
integration variable. We start with the single-particle differential phase-space element
in Eq.~\eqref{eq:n_particle_ps}. It can be written in the form
\begin{equation}\label{eq:lund_parametrization}
    \frac{{\rm d}^3\vec{p}_i}{(2\pi)^3\,2E_i}
    =\frac{1}{16\pi^2}\,{\rm d}p_{i,\perp}^2\,{\rm d}y_i\,\frac{\rm d\phi_i}{2\pi}\;,
\end{equation}
where $p_{i,\perp}$, $y_i$ and $\phi_i$ are the transverse momentum, rapidity 
and azimuthal angle of momentum $i$ in the laboratory frame, respectively. Many experimental 
analyses at hadron colliders require cuts on the transverse momentum and rapidity
of jets and other analysis objects, which are easily implemented in this parametrization,
leading to an excellent efficiency of the integration algorithm. 

The remaining task is to implement the delta function in Eq.~\eqref{eq:n_particle_ps}.
This is achieved by combining the integral over one of the momenta, say $p_n$,
with the integration over the light-cone momentum fractions used to convolute
the partonic cross section with the PDFs. We obtain
\begin{equation}
  \begin{split}
  {\rm d}x_a{\rm d}x_b\,{\rm d}\Phi_n(a,b;1,\ldots,n)
  =&\;\frac{{\rm d}P_+{\rm d}P_-}{s}\,
    \left[\,\prod_{i=1}^{n-1}\frac{1}{16\pi^2}\,
      {\rm d}p_{i,\perp}^2\,{\rm d}y_i\,\frac{\rm d\phi_i}{2\pi}\,\right]\\
  &\;\times\frac{{\rm d}^4p_n}{(2\pi)^3}\,\delta(p_n^2-s_n)\Theta(E_n)\;
    (2\pi)^4\delta^{(4)}\bigg(p_a+p_b-\sum_{i=1}^{n-1} p_i-p_n\bigg)\;,    
  \end{split}
\end{equation}
where $s$ is the hadronic center-of-mass energy, and $P_\pm=P_0\pm P_z$ is defined
using $P=\sum_{i=1}^{n-1}p_i$.
Changing the integration variables from $P_+$ and $P_-$ to $s_n$ and $y_n$,
it is straightforward to evaluate the delta functions, and we obtain the final expression
\begin{equation}\label{eq:t-channel_ps}
  \begin{split}
  {\rm d}x_a{\rm d}x_b\,{\rm d}\Phi_n(a,b;1,\ldots,n)
  =&\;\frac{2\pi}{s}\left[\,\prod_{i=1}^{n-1}\frac{1}{16\pi^2}\,
    {\rm d}p_{i,\perp}^2\,{\rm d}y_i\,\frac{\rm d\phi_i}{2\pi}\,\right]\,{\rm d}y_n\;.    
  \end{split}
\end{equation}
This form of the differential phase-space element is particularly suited for the
production of electroweak vector bosons ($W$, $Z$ and $\gamma$) in association with
any number of jets. However, it may not be optimal for phase-space generation
when there are strong hierarchies in transverse momenta of the jets, that may be
better described by phase-space mappings similar to Fig.~\ref{fig:factorization_example}~(c).

The differential decay phase-space elements occurring in Fig.~\ref{fig:factorization_example}~(b)
and~(c) are easily composed from the corresponding expressions for two-body decays. In the frame
of a time-like momentum $P$, this differential phase-space element can be written as
\begin{equation}\label{eq:two_body_ps}
    {\rm d}\Phi_{2}(\{1,2\};1,2)=
    \frac{1}{16\pi^2}\frac{\sqrt{(p_1P)^2-p_1^2P^2}^{\,3}}{((p_1P)(p_1p_2)-p_1^2(p_2P))P^2}\,
    {\rm d}\cos\theta_1^{(P)}{\rm d}\phi_1^{(P)}\;.
\end{equation}
Typically, this is evaluated in the center-of-mass frame of the combined momentum, 
$p_1+p_2$, where it simplifies to
\begin{equation}\label{eq:s-channel_ps}
  {\rm d}\Phi_2(\{1,2\};1,2)
  =\frac{1}{16\pi^2}\frac{\sqrt{(p_1p_2)^2-p_1^2p_2^2}}{(p_1+p_2)^2}\,
     {\rm d}\cos\theta_1^{\{1,2\}}\,{\rm d}\phi_1^{\{1,2\}}\;.
\end{equation}
Equations~\eqref{eq:t-channel_ps} and~\eqref{eq:s-channel_ps} form the basic building
blocks of our algorithm.

\subsection{The multi-channel}
\label{sec:multi_channel}
An optimal integrator for a particular squared Feynman diagram would be
composed of a combination of the t-channel map in Eq.~\eqref{eq:t-channel_ps}
and potentially a number of s-channel maps in Eq.~\eqref{eq:s-channel_ps},
as sketched for various configurations in Fig.~\ref{fig:factorization_example}.
The complete integrand will almost never consist of a single Feynman diagram squared,
and it is therefore more appropriate to combine various such integrators in order
to map out different structures in the full integrand.\footnote{An alternative option
  is to partition the integrand into terms which exhibit the structure of an
  individual diagram~\cite{Maltoni:2002qb}.} 
Each of those mappings is conventionally called a phase-space ``channel'', 
and each channel is a valid phase-space integrator in it's own right. 
They can be combined using the multi-channel technique, which was introduced
in~\cite{Kleiss:1994qy}. We refer the reader to the original publication for
the details of this method. Here we will briefly describe how the individual 
channels are constructed in our integrator.

We begin by extracting the three-particle vertices from the interaction model.
Given a set of external flavors, we can use the vertex information to construct
all possible topologies of Feynman diagrams with the maximum number of propagators.
For each topology, we apply the following algorithm:
If an s-channel propagator is found, we use the factorization formula,
Eq.~\eqref{eq:split_ps} to split the differential phase-space element into 
a production and a decay part. This procedure starts with the external states
and it is repeated until no more factorization is possible.
As the number of possible s-channel topologies grows factorially 
in many cases, our algorithm provides an option to limit the maximum number of
s-channels that are implemented. 
This helps to tailor the integrator to the problem at hand and allows to control the computational complexity.
Throughout the paper, we will refer to including the maximum number of s-channels as \Chili and
limiting the results to the minimum number of allowed s-channels (1 for $W$ and $Z$ processes and 0 otherwise) as
\Chili (basic).

Following standard practice, we generate the virtuality of the intermediate 
s-channel pseudo-particles using a Breit-Wigner distribution if the particle
has a mass and width, or following a ${\rm d}s/s^\alpha$ distribution ($\alpha<1$),
if the particle is massless.
The transverse momenta in Eq.~\eqref{eq:t-channel_ps}
are generated according to ${\rm d}p_\perp^2/(2p_{\perp,c}+p_\perp)^2$,
where
$p_{\perp,c}$ is an adjustable parameter that can be used to maximize efficiency,
e.g.\ by setting it to the jet transverse momentum cut. The rapidities in 
Eq.~\eqref{eq:t-channel_ps} and the angles in Eq.~\eqref{eq:s-channel_ps}
are generated using a flat prior distribution. 
The virtuality ($s$) for an intermediate resonance following a
Breit-Wigner distribution can be generated for a particle of mass $M$
and width $\Gamma$ for an invariant mass squared between $s_{\rm min}$ 
and $s_{\rm max}$ with random number $r \in [0,1)$ by
\begin{equation}
    s = M^2 + M \Gamma \tan\left(y_{\rm min} + r \left(y_{\rm max} - y_{\rm min}\right)\right)\;,
\end{equation}
where we have defined $y_{\rm min,max} = \arctan\left[\left(s_{\rm max,min} - M^2\right)/\left(M \Gamma\right)\right]$.

\subsection{Next-to-leading order calculations and dipole mappings}
\label{sec:dipole_mapping}
The integration of real-emission corrections in next-to-leading order QCD or QED
calculations poses additional challenges for a phase-space integration algorithm.
In order to achieve a local cancellation of singularities, subtraction methods are
typically employed in these calculations~\cite{Frixione:1995ms,Catani:1996vz}.
This makes the behavior of the integrand less predictable than at leading order,
and therefore complicates the construction of integration channels. Various approaches 
have been devised to deal with the problem. We adopt a solution that is based on the 
on-shell momentum mapping technique used in the Catani-Seymour dipole subtraction
scheme~\cite{Catani:1996vz,Catani:2002hc} and that has long been used in generators
such as MCFM~\cite{Ellis:2009zw,Campbell:2011bn,Campbell:2019dru} and 
MUNICH~\cite{Grazzini:2017mhc}.\footnote{We make this feature available only for use within
  \Sherpa, but a future version of our stand-alone code will support it as well.}

Following Ref.~\cite{Catani:1996vz}, there are four different types of local
infrared subtraction terms that are used to make real-emission corrections and
virtual corrections in NLO calculations separately infrared finite. They are 
classified according to the type of collinear divergence (initial state or final state)
and the type of color spectator parton (initial state or final state). 
The massless on-shell phase-space mapping for the final-final configuration (FF) reads
\begin{equation}\label{eq:cs_mappings_ff}
  {\rm d}\Phi_n^{\rm(FF)}(a,b;1,\ldots,n)=
  {\rm d}\Phi_{n-1}(a,b;1,\ldots,\widetilde{\imath\jmath},\ldots,\tilde{k},\ldots,n)\,
  \frac{2\tilde{p}_{ij}\tilde{p}_k}{16\pi^2}\,{\rm d}y_{ij,k}{\rm d}\tilde{z}_i\,
  \frac{{\rm d}\phi}{2\pi}\,(1-y_{ij,k})\;.
\end{equation}
where
\begin{equation}
  p_i^\mu=\tilde{z}_i\,\tilde{p}_{ij}^\mu+(1-\tilde{z}_i)\,y_{ij,k}\,\tilde{p}_k^\mu+k_\perp^\mu\;,\qquad
  p_k^\mu=(1-y_{ij,k})\,\tilde{p}_k^\mu\;,\qquad
  p_j^\mu=\tilde{p}_{ij}+\tilde{p}_k-p_i-p_k\;,
\end{equation}
and where $k_\perp^2=-\tilde{z}_i(1-\tilde{z}_i)y_{ij,k}\,2\tilde{p}_{ij}\tilde{p}_k$ is determined
by the on-shell conditions.\\
The massless on-shell phase-space mapping for the final-initial and initial-final 
configurations (FI/IF) reads
\begin{equation}\label{eq:cs_mappings_fi}
  {\rm d}\Phi_n^{\rm(FI/IF)}(a,b;1,\ldots,n)=
  {\rm d}\Phi_{n-1}(\tilde{a},b;1,\ldots,\widetilde{\imath\jmath},\ldots,n)\,
  \frac{2\tilde{p}_{ij}p_a}{16\pi^2}\,{\rm d}\tilde{z}_i{\rm d}x_{ij,a}\,
  \frac{{\rm d}\phi}{2\pi}\;.
\end{equation}
where
\begin{equation}
  p_i^\mu=\tilde{z}_i\,\tilde{p}_{ij}^\mu+(1-\tilde{z}_i)\,\frac{1-x_{ij,a}}{x_{ij,a}}\,\,\tilde{p}_a^\mu+k_\perp^\mu\;,\qquad
  p_a^\mu=\frac{1}{x_{ij,a}}\,\tilde{p}_a^\mu\;,\qquad
  p_j^\mu=\tilde{p}_{ij}-\tilde{p}_a+\tilde{p}_a-\tilde{p}_i\;,
\end{equation}
and where $k_\perp^2=-\tilde{z}_i(1-\tilde{z}_i)(1-x_{ij,a})/x_{ij,a}\,2\tilde{p}_{ij}\tilde{p}_a$.\\
The massless on-shell phase-space mapping for the initial-initial configurations (II) reads
\begin{equation}\label{eq:cs_mappings_ii}
  {\rm d}\Phi_n^{\rm(II)}(a,b;1,\ldots,n)=
  {\rm d}\Phi_{n-1}(\widetilde{a\imath},b;\tilde{1},\ldots,\tilde{n})\,
  \frac{2p_ap_b}{16\pi^2}\,{\rm d}\tilde{v}_i{\rm d}x_{i,ab}\,
  \frac{{\rm d}\phi}{2\pi}\;.
\end{equation}
where
\begin{equation}
  p_i^\mu=\frac{1-x_{i,ab}-\tilde{v}_i}{x_{i,ab}}\,\tilde{p}_{a}^\mu+\tilde{v}_i\,p_b^\mu+k_\perp^\mu\;,\qquad
  p_a^\mu=\frac{1}{x_{i,ab}}\,\tilde{p}_{ai}^\mu\;,\qquad
  p_j^\mu=\Lambda^\mu_{\;\nu}(K,\tilde{K})\tilde{p}_{j}^\nu\quad\forall j\in\{1,\ldots,n\},j\neq i\;,
\end{equation}
and where $k_\perp^2=-(1-x_{i,ab}-\tilde{v})/x_{ij,a}\,\tilde{v}_i\,2\tilde{p}_{ai}p_b$.
The transformation, $\Lambda^\mu_{\;\nu}(K,\tilde{K})$, is defined in Sec.~5.5 of Ref.~\cite{Catani:1996vz}.
The three above mappings are sufficient to treat any real-emission correction in massless QCD.
We infer the possible dipole configurations from the flavor structure of the process and combine
all possible mappings into a multi-channel integrator~\cite{Kleiss:1994qy}. 

\subsection{Combination with normalizing-flow based integrators}
\label{sec:iflow}
With the development of modern machine learning methods, new techniques for adaptive
Monte-Carlo integration have emerged, which are based on the extension~\cite{1808.03856,durkan2019neural}
of a nonlinear independent components estimation technique~\cite{1410.8516,1605.08803},
also known as a normalizing flow. They have been used to develop integration algorithms based
on existing multi-channel approaches~\cite{Bothmann:2020ywa,Gao:2020zvv,Heimel:2022wyj,Butter:2022rso}.
One of the main obstacles to scaling such approaches to high multiplicity has been the fact
that the underlying phase-space mappings are
used as individual mappings in a multi-channel phase-space generator.
The channel selection requires additional hyperparameters,
which increases the dimensionality of the optimization problem. Here we propose a different strategy.
We observe that the basic t-channel integration algorithm implementing Eq.~\eqref{eq:t-channel_ps}
requires the minimal amount of random numbers, and shows a good efficiency (cf.\ Sec.~\ref{sec:performance}).
It is therefore ideally suited to provide a basic mapping of the $n$-particle phase space at hadron colliders
into a $3n-4+2$ dimensional unit hypercube, required for combination with normalizing-flow based integrators.
We provide Python bindings in \Chili via nanobind~\cite{nanobind} and a dedicated 
Tensorflow~\cite{Abadi_TensorFlow_Large-scale_machine_2015} interface. This allows the use of the 
iFlow~\cite{Gao:2020vdv} and \MadNIS~\cite{Heimel:2022wyj} frameworks to test this idea,
and to evaluate the performance of this novel algorithm. 

\section{Performance Benchmarks}
\label{sec:performance}
\begin{table}[t]
  \begin{tabularx}{0.4925\textwidth}{>{\raggedright\arraybackslash}X|
    >{\centering\arraybackslash}X|>{\centering\arraybackslash}X| 
    >{\centering\arraybackslash}X|>{\centering\arraybackslash}X| 
    >{\centering\arraybackslash}X|>{\centering\arraybackslash}X}
       Process &  \multicolumn{2}{c|}{\Sherpa} & \multicolumn{2}{c|}{\Chili} &
       \multicolumn{2}{c}{\Chili (basic)} \\
       \; & $\Delta\sigma/\sigma$ & $\eta$ & $\Delta\sigma/\sigma$ &
       $\eta$ & $\Delta\sigma/\sigma$ & $\eta$ \\[-1mm]
       \; & {\scriptsize 6M pts} & {\scriptsize 100 evts} & {\scriptsize 6M pts} &
       {\scriptsize 100 evts} & {\scriptsize 6M pts} & {\scriptsize 100 evts} \\\hline 
       $W^+$+1j & 0.5\textperthousand & $7\times 10^{-2}$ & 0.6\textperthousand & $9\times 10^{-2}$ & 0.6\textperthousand & $9\times 10^{-2}$ \\
       $W^+$+2j & 1.2\textperthousand & $9\times 10^{-3}$ & 1.1\textperthousand & $2\times 10^{-2}$ & 1.2\textperthousand & $1\times 10^{-2}$ \\
       $W^+$+3j & 2.0\textperthousand & $1\times 10^{-3}$ & 2.0\textperthousand & $4\times 10^{-3}$ & 2.9\textperthousand & $2\times 10^{-3}$ \\
       $W^+$+4j & 3.7\textperthousand & $2\times 10^{-4}$ & 4.9\textperthousand & $7\times 10^{-4}$ & 6.0\textperthousand & $3\times 10^{-4}$ \\
       $W^+$+5j & 7.2\textperthousand & $4\times 10^{-5}$ & 22\textperthousand & $1\times 10^{-5}$ & 26\textperthousand & $1\times 10^{-5}$ \\
  \end{tabularx}\hfill
  \begin{tabularx}{0.4925\textwidth}{>{\raggedright\arraybackslash}X|
    >{\centering\arraybackslash}X|>{\centering\arraybackslash}X| 
    >{\centering\arraybackslash}X|>{\centering\arraybackslash}X| 
    >{\centering\arraybackslash}X|>{\centering\arraybackslash}X}
       Process &  \multicolumn{2}{c|}{\Sherpa} & \multicolumn{2}{c|}{\Chili} &
       \multicolumn{2}{c}{\Chili (basic)} \\
       \; & $\Delta\sigma/\sigma$ & $\eta$ & $\Delta\sigma/\sigma$ &
       $\eta$ & $\Delta\sigma/\sigma$ & $\eta$ \\[-1mm]
       \; & {\scriptsize 6M pts} & {\scriptsize 100 evts} & {\scriptsize 6M pts} &
       {\scriptsize 100 evts} & {\scriptsize 6M pts} & {\scriptsize 100 evts} \\\hline 
       $Z$+1j & 0.4\textperthousand & $2\times 10^{-1}$ & 0.5\textperthousand & $1\times 10^{-1}$ & 0.5\textperthousand & $1\times 10^{-1}$ \\
       $Z$+2j & 0.8\textperthousand & $2\times 10^{-2}$ & 0.8\textperthousand & $3\times 10^{-2}$ & 1.0\textperthousand & $2\times 10^{-2}$ \\
       $Z$+3j & 1.3\textperthousand & $4\times 10^{-3}$ & 1.6\textperthousand & $7\times 10^{-3}$ & 2.5\textperthousand & $4\times 10^{-3}$ \\
       $Z$+4j & 2.2\textperthousand & $8\times 10^{-4}$ & 3.6\textperthousand & $1\times 10^{-3}$ & 5.0\textperthousand & $6\times 10^{-4}$ \\
       $Z$+5j & 3.7\textperthousand & $1\times 10^{-4}$ & 11\textperthousand & $1\times 10^{-4}$ & 13\textperthousand & $2\times 10^{-4}$ \\
  \end{tabularx}\\\vskip 3mm
  \begin{tabularx}{0.4925\textwidth}{>{\raggedright\arraybackslash}X|
    >{\centering\arraybackslash}X|>{\centering\arraybackslash}X| 
    >{\centering\arraybackslash}X|>{\centering\arraybackslash}X| 
    >{\centering\arraybackslash}X|>{\centering\arraybackslash}X}
       Process &  \multicolumn{2}{c|}{\Sherpa} & \multicolumn{2}{c|}{\Chili} &
       \multicolumn{2}{c}{\Chili (basic)} \\
       \; & $\Delta\sigma/\sigma$ & $\eta$ & $\Delta\sigma/\sigma$ &
       $\eta$ & $\Delta\sigma/\sigma$ & $\eta$ \\[-1mm]
       \; & {\scriptsize 6M pts} & {\scriptsize 100 evts} & {\scriptsize 6M pts} &
       {\scriptsize 100 evts} & {\scriptsize 6M pts} & {\scriptsize 100 evts} \\\hline 
       $h$+1j & 0.4\textperthousand & $2\times10^{-1}$ & 0.4\textperthousand & $2\times10^{-1}$ & 0.4\textperthousand & $2\times10^{-1}$ \\
       $h$+2j & 0.8\textperthousand & $2\times10^{-2}$ & 0.6\textperthousand & $5\times10^{-2}$ & 0.6\textperthousand & $5\times10^{-2}$ \\
       $h$+3j & 1.4\textperthousand & $3\times10^{-3}$ & 0.9\textperthousand & $2\times10^{-2}$ & 0.9\textperthousand & $2\times10^{-2}$ \\
       $h$+4j & 2.4\textperthousand & $6\times10^{-4}$ & 1.6\textperthousand & $6\times10^{-3}$ & 1.7\textperthousand & $7\times10^{-3}$ \\
       $h$+5j & 4.5\textperthousand & $1\times10^{-4}$ & 3.2\textperthousand & $1\times10^{-3}$ & 3.6\textperthousand & $1\times10^{-3}$ \\
  \end{tabularx}\hfill
  \begin{tabularx}{0.4925\textwidth}{>{\raggedright\arraybackslash}X|
    >{\centering\arraybackslash}X|>{\centering\arraybackslash}X| 
    >{\centering\arraybackslash}X|>{\centering\arraybackslash}X| 
    >{\centering\arraybackslash}X|>{\centering\arraybackslash}X}
       Process &  \multicolumn{2}{c|}{\Sherpa} & \multicolumn{2}{c|}{\Chili} &
       \multicolumn{2}{c}{\Chili (basic)} \\
       \; & $\Delta\sigma/\sigma$ & $\eta$ & $\Delta\sigma/\sigma$ &
       $\eta$ & $\Delta\sigma/\sigma$ & $\eta$ \\[-1mm]
       \; & {\scriptsize 6M pts} & {\scriptsize 100 evts} & {\scriptsize 6M pts} &
       {\scriptsize 100 evts} & {\scriptsize 6M pts} & {\scriptsize 100 evts} \\\hline 
       $t\bar{t}$+0j & 0.6\textperthousand & $1\times 10^{-1}$ & 0.6\textperthousand & $1\times 10^{-1}$ & 0.6\textperthousand & $1\times 10^{-1}$ \\
       $t\bar{t}$+1j & 0.9\textperthousand & $2\times 10^{-2}$ & 0.6\textperthousand & $6\times 10^{-2}$ & 0.9\textperthousand & $3\times 10^{-2}$ \\
       $t\bar{t}$+2j & 1.4\textperthousand & $4\times 10^{-3}$ & 0.9\textperthousand & $2\times 10^{-2}$ & 1.4\textperthousand & $1\times 10^{-2}$ \\
       $t\bar{t}$+3j & 2.6\textperthousand & $7\times 10^{-4}$ & 1.5\textperthousand & $7\times 10^{-3}$ & 2.9\textperthousand & $2\times 10^{-3}$ \\
       $t\bar{t}$+4j & 4.0\textperthousand & $1\times 10^{-4}$ & 3.2\textperthousand & $1\times 10^{-3}$ & 3.5\textperthousand & $8\times 10^{-4}$ \\
  \end{tabularx}\\\vskip 3mm
  \begin{tabularx}{0.4925\textwidth}{>{\raggedright\arraybackslash}X|
    >{\centering\arraybackslash}X|>{\centering\arraybackslash}X| 
    >{\centering\arraybackslash}X|>{\centering\arraybackslash}X| 
    >{\centering\arraybackslash}X|>{\centering\arraybackslash}X}
       Process &  \multicolumn{2}{c|}{\Sherpa} & \multicolumn{2}{c|}{\Chili} &
       \multicolumn{2}{c}{\Chili (basic)} \\
       \; & $\Delta\sigma/\sigma$ & $\eta$ & $\Delta\sigma/\sigma$ &
       $\eta$ & $\Delta\sigma/\sigma$ & $\eta$ \\[-1mm]
       \; & {\scriptsize 6M pts} & {\scriptsize 100 evts} & {\scriptsize 6M pts} &
       {\scriptsize 100 evts} & {\scriptsize 6M pts} & {\scriptsize 100 evts} \\\hline 
       $\gamma$+1j & 0.4\textperthousand & $2\times 10^{-1}$ & 0.6\textperthousand & $1\times 10^{-1}$ & 0.6\textperthousand & $1\times 10^{-1}$ \\
       $\gamma$+2j & 1.1\textperthousand & $7\times 10^{-3}$ & 2.2\textperthousand & $3\times 10^{-3}$ & 3.7\textperthousand & $1\times 10^{-3}$ \\
       $\gamma$+3j & 2.4\textperthousand & $5\times 10^{-4}$ & 4.9\textperthousand & $4\times 10^{-4}$ & 10\textperthousand & $1\times 10^{-4}$ \\
       $\gamma$+4j & 5.0\textperthousand & $7\times 10^{-5}$ & 20\textperthousand & $3\times 10^{-5}$ & 30\textperthousand & $4\times 10^{-5}$ \\
       $\gamma$+5j & 9.3\textperthousand & $2\times 10^{-5}$ & 28\textperthousand & $7\times 10^{-6}$ & 36\textperthousand & $2\times 10^{-6}$ \\
  \end{tabularx}\hfill
  \begin{tabularx}{0.4925\textwidth}{>{\raggedright\arraybackslash}X|
    >{\centering\arraybackslash}X|>{\centering\arraybackslash}X| 
    >{\centering\arraybackslash}X|>{\centering\arraybackslash}X| 
    >{\centering\arraybackslash}X|>{\centering\arraybackslash}X}
       Process &  \multicolumn{2}{c|}{\Sherpa} & \multicolumn{2}{c|}{\Chili} &
       \multicolumn{2}{c}{\Chili (basic)} \\
       \; & $\Delta\sigma/\sigma$ & $\eta$ & $\Delta\sigma/\sigma$ &
       $\eta$ & $\Delta\sigma/\sigma$ & $\eta$ \\[-1mm]
       \; & {\scriptsize 6M pts} & {\scriptsize 100 evts} & {\scriptsize 6M pts} &
       {\scriptsize 100 evts} & {\scriptsize 6M pts} & {\scriptsize 100 evts} \\\hline 
       2jets & 0.6\textperthousand & $5\times 10^{-2}$ & 0.4\textperthousand & $1\times 10^{-1}$ & 0.5\textperthousand & $7\times 10^{-2}$ \\
       3jets & 1.2\textperthousand & $5\times 10^{-3}$ & 1.0\textperthousand & $1\times 10^{-2}$ & 1.8\textperthousand & $7\times 10^{-3}$ \\
       4jets & 2.5\textperthousand & $5\times 10^{-4}$ & 2.0\textperthousand & $3\times 10^{-3}$ & 3.4\textperthousand & $1\times 10^{-3}$ \\
       5jets & 4.7\textperthousand & $9\times 10^{-5}$ & 5.1\textperthousand & $6\times 10^{-4}$ & 8.1\textperthousand & $2\times 10^{-4}$ \\
       6jets & 7.0\textperthousand & $2\times 10^{-5}$ & 15\textperthousand & $5\times 10^{-5}$ & 14\textperthousand & $4\times 10^{-5}$ \\
  \end{tabularx}\\
  \caption{Relative Monte-Carlo uncertainties, $\Delta\sigma/\sigma$, and unweighting efficiencies,
    $\eta$, in leading-order calculations. The center-of-mass energy is $\sqrt{s}=14$~TeV, jets are defined
    using the anti-$k_T$ algorithm with $p_{\perp,j}=30$~GeV and $|y_j|\le 6$. Vegas grids and multi-channel
    weights have been adapted using 1.2M non-zero phase-space points. For details see the main text.}
  \label{tab:comparison_democratic}
\end{table}
\begin{table}
  \begin{tabularx}{0.4925\textwidth}{>{\raggedright\arraybackslash}X|
    >{\centering\arraybackslash}X|>{\centering\arraybackslash}X| 
    >{\centering\arraybackslash}X|>{\centering\arraybackslash}X| 
    >{\centering\arraybackslash}X|>{\centering\arraybackslash}X}
       Process &  \multicolumn{2}{c|}{\Sherpa} & \multicolumn{2}{c|}{\Chili} &
       \multicolumn{2}{c}{\Chili (basic)} \\
       \; & $\Delta\sigma/\sigma$ & $\eta$ & $\Delta\sigma/\sigma$ &
       $\eta$ & $\Delta\sigma/\sigma$ & $\eta$ \\[-1mm]
       boosted & {\scriptsize 6M pts} & {\scriptsize 100 evts} & {\scriptsize 6M pts} &
       {\scriptsize 100 evts} & {\scriptsize 6M pts} & {\scriptsize 100 evts} \\\hline 
       $W^+$+2j & 1.4\textperthousand & $4\times10^{-3}$ & 1.4\textperthousand & $8\times10^{-3}$ & 2.5\textperthousand & $2\times10^{-3}$ \\
       $W^+$+3j & 2.5\textperthousand & $9\times10^{-4}$ & 3.8\textperthousand & $6\times10^{-4}$ & 6.9\textperthousand & $2\times10^{-4}$ \\
       $W^+$+4j & 4.2\textperthousand & $2\times10^{-4}$ & 10\textperthousand & $7\times10^{-5}$ & 17\textperthousand & $4\times10^{-5}$ \\
       $W^+$+5j & 7.2\textperthousand & $4\times10^{-5}$ & 27\textperthousand & $3\times10^{-6}$ & 48\textperthousand & $4\times10^{-6}$ \\
  \end{tabularx}\hfill
  \begin{tabularx}{0.4925\textwidth}{>{\raggedright\arraybackslash}X|
    >{\centering\arraybackslash}X|>{\centering\arraybackslash}X| 
    >{\centering\arraybackslash}X|>{\centering\arraybackslash}X| 
    >{\centering\arraybackslash}X|>{\centering\arraybackslash}X}
       Process &  \multicolumn{2}{c|}{\Sherpa} & \multicolumn{2}{c|}{\Chili} &
       \multicolumn{2}{c}{\Chili (basic)} \\
       \; & $\Delta\sigma/\sigma$ & $\eta$ & $\Delta\sigma/\sigma$ &
       $\eta$ & $\Delta\sigma/\sigma$ & $\eta$ \\[-1mm]
       boosted & {\scriptsize 6M pts} & {\scriptsize 100 evts} & {\scriptsize 6M pts} &
       {\scriptsize 100 evts} & {\scriptsize 6M pts} & {\scriptsize 100 evts} \\\hline 
       $Z$+2j & 1.0\textperthousand & $9\times10^{-3}$ & 1.1\textperthousand & $1\times10^{-2}$ & 1.8\textperthousand & $6\times10^{-3}$ \\
       $Z$+3j & 1.6\textperthousand & $2\times10^{-3}$ & 2.5\textperthousand & $2\times10^{-3}$ & 5.0\textperthousand & $5\times10^{-4}$ \\
       $Z$+4j & 2.8\textperthousand & $4\times10^{-4}$ & 7.6\textperthousand & $2\times10^{-4}$ & 27\textperthousand & $6\times10^{-5}$ \\
       $Z$+5j & 4.6\textperthousand & $9\times10^{-5}$ & 15\textperthousand & $3\times10^{-5}$ & 33\textperthousand & $2\times10^{-5}$
  \end{tabularx}\\\vskip 3mm
  \begin{tabularx}{0.4925\textwidth}{>{\raggedright\arraybackslash}X|
    >{\centering\arraybackslash}X|>{\centering\arraybackslash}X| 
    >{\centering\arraybackslash}X|>{\centering\arraybackslash}X| 
    >{\centering\arraybackslash}X|>{\centering\arraybackslash}X}
       Process &  \multicolumn{2}{c|}{\Sherpa} & \multicolumn{2}{c|}{\Chili} &
       \multicolumn{2}{c}{\Chili (basic)} \\
       \; & $\Delta\sigma/\sigma$ & $\eta$ & $\Delta\sigma/\sigma$ &
       $\eta$ & $\Delta\sigma/\sigma$ & $\eta$ \\[-1mm]
       boosted & {\scriptsize 6M pts} & {\scriptsize 100 evts} & {\scriptsize 6M pts} &
       {\scriptsize 100 evts} & {\scriptsize 6M pts} & {\scriptsize 100 evts} \\\hline 
       $h$+2j & 1.1\textperthousand & $8\times10^{-3}$ & 0.7\textperthousand & $4\times10^{-2}$ & 0.7\textperthousand & $3\times10^{-2}$\\
       $h$+3j & 1.8\textperthousand & $2\times10^{-3}$ & 1.0\textperthousand & $1\times10^{-2}$ & 1.1\textperthousand & $1\times10^{-2}$\\
       $h$+4j & 3.0\textperthousand & $4\times10^{-4}$ & 1.7\textperthousand & $3\times10^{-3}$ & 1.6\textperthousand & $4\times10^{-3}$\\
       $h$+5j & 4.8\textperthousand & $9\times10^{-5}$ & 4.2\textperthousand & $7\times10^{-4}$ & 3.1\textperthousand & $1\times10^{-4}$\\
  \end{tabularx}\hfill
  \begin{tabularx}{0.4925\textwidth}{>{\raggedright\arraybackslash}X|
    >{\centering\arraybackslash}X|>{\centering\arraybackslash}X| 
    >{\centering\arraybackslash}X|>{\centering\arraybackslash}X| 
    >{\centering\arraybackslash}X|>{\centering\arraybackslash}X}
       Process &  \multicolumn{2}{c|}{\Sherpa} & \multicolumn{2}{c|}{\Chili} &
       \multicolumn{2}{c}{\Chili (basic)} \\
       \; & $\Delta\sigma/\sigma$ & $\eta$ & $\Delta\sigma/\sigma$ &
       $\eta$ & $\Delta\sigma/\sigma$ & $\eta$ \\[-1mm]
       boosted & {\scriptsize 6M pts} & {\scriptsize 100 evts} & {\scriptsize 6M pts} &
       {\scriptsize 100 evts} & {\scriptsize 6M pts} & {\scriptsize 100 evts} \\\hline 
       $\gamma$+2j & 1.4\textperthousand & $4\times10^{-3}$ & 2.3\textperthousand & $2\times10^{-3}$ & 2.3\textperthousand & $2\times10^{-3}$ \\
       $\gamma$+3j & 2.3\textperthousand & $7\times10^{-4}$ & 4.3\textperthousand & $4\times10^{-4}$ & 9.0\textperthousand & $1\times10^{-4}$ \\
       $\gamma$+4j & 4.0\textperthousand & $2\times10^{-4}$ & 9.9\textperthousand & $1\times10^{-4}$ & 25\textperthousand & $1\times10^{-5}$ \\
       $\gamma$+5j & 7.3\textperthousand & $2\times10^{-5}$ & 36\textperthousand & $1\times10^{-6}$ & 49\textperthousand & $3\times10^{-6}$ \\
  \end{tabularx}\\\vskip 3mm
  \begin{tabularx}{0.4925\textwidth}{>{\raggedright\arraybackslash}X|
    >{\centering\arraybackslash}X|>{\centering\arraybackslash}X| 
    >{\centering\arraybackslash}X|>{\centering\arraybackslash}X| 
    >{\centering\arraybackslash}X|>{\centering\arraybackslash}X}
       Process &  \multicolumn{2}{c|}{\Sherpa} & \multicolumn{2}{c|}{\Chili} &
       \multicolumn{2}{c}{\Chili (basic)} \\
       \; & $\Delta\sigma/\sigma$ & $\eta$ & $\Delta\sigma/\sigma$ &
       $\eta$ & $\Delta\sigma/\sigma$ & $\eta$ \\[-1mm]
       boosted & {\scriptsize 6M pts} & {\scriptsize 100 evts} & {\scriptsize 6M pts} &
       {\scriptsize 100 evts} & {\scriptsize 6M pts} & {\scriptsize 100 evts} \\\hline 
       $t\bar{t}$+1j & 1.0\textperthousand & $1\times10^{-2}$ & 0.7\textperthousand & $4\times10^{-2}$ & 1.5\textperthousand & $1\times10^{-2}$ \\
       $t\bar{t}$+2j & 2.0\textperthousand & $1\times10^{-3}$ & 1.1\textperthousand & $1\times10^{-2}$ & 2.3\textperthousand & $2\times10^{-3}$ \\
       $t\bar{t}$+3j & 3.2\textperthousand & $4\times10^{-4}$ & 1.9\textperthousand & $3\times10^{-3}$ & 3.7\textperthousand & $8\times10^{-4}$ \\
       $t\bar{t}$+4j & 4.9\textperthousand & $1\times10^{-4}$ & 3.8\textperthousand & $7\times10^{-4}$ & 8.4\textperthousand & $2\times10^{-4}$ \\
  \end{tabularx}\hfill
  \begin{tabularx}{0.4925\textwidth}{>{\raggedright\arraybackslash}X|
    >{\centering\arraybackslash}X|>{\centering\arraybackslash}X| 
    >{\centering\arraybackslash}X|>{\centering\arraybackslash}X| 
    >{\centering\arraybackslash}X|>{\centering\arraybackslash}X}
       Process &  \multicolumn{2}{c|}{\Sherpa} & \multicolumn{2}{c|}{\Chili} &
       \multicolumn{2}{c}{\Chili (basic)} \\
       \; & $\Delta\sigma/\sigma$ & $\eta$ & $\Delta\sigma/\sigma$ &
       $\eta$ & $\Delta\sigma/\sigma$ & $\eta$ \\[-1mm]
       $m_{jj}$ cut & {\scriptsize 6M pts} & {\scriptsize 100 evts} & {\scriptsize 6M pts} &
       {\scriptsize 100 evts} & {\scriptsize 6M pts} & {\scriptsize 100 evts} \\\hline 
       $h$+2j & 0.9\textperthousand & $1\times10^{-2}$ & 0.8\textperthousand & $1\times10^{-2}$ & 0.9\textperthousand & $1\times10^{-2}$ \\
       $h$+3j & 1.9\textperthousand & $1\times10^{-3}$ & 1.2\textperthousand & $5\times10^{-3}$ & 1.3\textperthousand & $4\times10^{-3}$ \\
       $h$+4j & 4.1\textperthousand & $2\times10^{-4}$ & 1.8\textperthousand & $2\times10^{-3}$ & 2.3\textperthousand & $1\times10^{-3}$ \\
       $h$+5j & 16\textperthousand & $5\times10^{-5}$ & 5.0\textperthousand & $2\times10^{-4}$ & 4.5\textperthousand & $5\times10^{-4}$ \\
  \end{tabularx}\\
  \caption{Relative Monte-Carlo uncertainties, $\Delta\sigma/\sigma$, and unweighting efficiencies,
    $\eta$, in leading-order calculations for boosted event topologies. The center-of-mass energy 
    is $\sqrt{s}=14$~TeV, jets are defined using the anti-$k_T$ algorithm with $p_{\perp,j}=30$~GeV and 
    $|y_j|\le 6$. We require a leading jet at $p_{\perp,j1}\ge300$~GeV. Vegas grids and multi-channel
    weights have been adapted using 1.2M non-zero phase-space points. For details see the main text.}
  \label{tab:comparison_boosted}
\end{table}
\begin{table}
  \begin{tabularx}{\textwidth}{l|>{\centering\arraybackslash}X|>{\centering\arraybackslash}X|
    >{\centering\arraybackslash}X|>{\centering\arraybackslash}Xp{1mm}
    l|>{\centering\arraybackslash}X|>{\centering\arraybackslash}X|
    >{\centering\arraybackslash}X|>{\centering\arraybackslash}X}
       Process &  \multicolumn{2}{c|}{\Sherpa} & \multicolumn{2}{c}{\Chili (basic)} &&
       Process &  \multicolumn{2}{c|}{\Sherpa} & \multicolumn{2}{c}{\Chili (basic)} \\
       1M pts &  $\Delta\sigma/\sigma$ & $\varepsilon_{\rm cut}$ &
       $\Delta\sigma/\sigma$ & $\varepsilon_{\rm cut}$ &&
       1M pts &  $\Delta\sigma/\sigma$ & $\varepsilon_{\rm cut}$ &
       $\Delta\sigma/\sigma$ & $\varepsilon_{\rm cut}$ \\\cline{1-5}\cline{7-11} 
       $W^+$+1j / B-like & 1.3\textperthousand & 43\% & 1.4\textperthousand & 99\% &&
       $h$+1j / B-like & 1.3\textperthousand & 56\% & 0.7\textperthousand & 99\% \\
       \hphantom{$W^+$+1j /} R-like & 4.1\textperthousand & 46\% & 3.6\textperthousand & 58\% &&
       \hphantom{$h$+1j /} R-like & 3.0\textperthousand & 52\% & 2.1\textperthousand & 69\% \\
       $W^+$+2j / B-like & 2.2\textperthousand & 37\% & 4.4\textperthousand & 99\% &&
       $h$+2j / B-like & 2.6\textperthousand & 34\% & 1.4\textperthousand & 99\% \\
       \hphantom{$W^+$+2j /} R-like & 1.4\% & 74\% & 1.5\% & 80\% &&
       \hphantom{$h$+2j /} R-like & 8.1\textperthousand & 68\% & 8.2\textperthousand & 87\% \\
       $W^+$+3j$^\dagger\!\!\!$ / B-like & 2.8\% & 33\% & 3.5\% & 97\% &&
       $h$+3j$^*\!\!\!$ / B-like & 2.3\% & 29\% & 1.0\% & 96\% \\
       \hphantom{$W^+$+3j /} R-like & 3.0\% & 75\% & 4.3\% & 87\% &&
       \hphantom{$h$+3j /} R-like & 2.0\% & 65\% & 2.0\% & 83\% \\
       \multicolumn{11}{c}{}\\
       Process &  \multicolumn{2}{c|}{\Sherpa} & \multicolumn{2}{c}{\Chili (basic)} &&
       Process &  \multicolumn{2}{c|}{\Sherpa} & \multicolumn{2}{c}{\Chili (basic)} \\
       1M pts &  $\Delta\sigma/\sigma$ & $\varepsilon_{\rm cut}$ &
       $\Delta\sigma/\sigma$ & $\varepsilon_{\rm cut}$ &&
       1M pts &  $\Delta\sigma/\sigma$ & $\varepsilon_{\rm cut}$ &
       $\Delta\sigma/\sigma$ & $\varepsilon_{\rm cut}$ \\\cline{1-5}\cline{7-11} 
       $t\bar{t}$+0j / B-like & 0.4\textperthousand & 99\% & 0.8\textperthousand & 99\% &&
       2jets / B-like & 1.5\textperthousand & 34\% & 0.7\textperthousand & 99\% \\
       \hphantom{$t\bar{t}$+0j /} R-like & 0.2\textperthousand & 99\% & 0.3\textperthousand & 99\% &&
       \hphantom{2jets /} R-like & 8.3\textperthousand & 76\% & 4.3\textperthousand & 89\% \\
       $t\bar{t}$+1j / B-like & 1.7\textperthousand & 61\% & 1.7\textperthousand & 99\% &&
       3jets / B-like & 4.2\% & 9.6\% & 6.1\textperthousand & 88\% \\
       \hphantom{$t\bar{t}$+1j /} R-like & 5.8\textperthousand & 82\% & 5.9\textperthousand & 92\% &&
       \hphantom{3jets /} R-like & 4.5\% & 56\% & 3.7\% & 81\% \\
       $t\bar{t}$+2j / B-like & 1.5\% & 45\% & 1.0\% & 98\% &&
       4jets$^*\!\!\!$ / B-like & 4.8\% & 12\% & 3.2\% & 90\% \\
       \hphantom{$t\bar{t}$+2j /} R-like & 1.4\% & 78\% & 1.7\% & 85\% &&
       \hphantom{4jets /} R-like & 4.7\% & 50\% & 3.7\% & 79\% \\
  \end{tabularx}
  \caption{Relative Monte-Carlo uncertainties, $\Delta\sigma/\sigma$, and cut efficiencies,
    $\varepsilon_{\rm cut}$, in next-to-leading order calculations. The center-of-mass energy is 
    $\sqrt{s}=14$~TeV, jets are defined using the anti-$k_T$ algorithm with $p_{\perp,j}=30$~GeV
    and $|y_j|\le 6$. 
    The superscript $^\dagger$ indicates a factor 10 reduction in the number of points
    to evaluate the Born-like components. The superscript $^*$ indicates a factor 10 reduction
    in the number of points to evaluate the Born-like components and the usage of a global
    $K$-factor as a stand-in for the finite virtual corrections.}
  \label{tab:comparison_nlo}
\end{table}
\begin{table}[t]
  \begin{tabularx}{0.4925\textwidth}{>{\raggedright\arraybackslash}X|
    >{\centering\arraybackslash}X|>{\centering\arraybackslash}X| 
    >{\centering\arraybackslash}X|>{\centering\arraybackslash}X| 
    >{\centering\arraybackslash}X|>{\centering\arraybackslash}X}
       Process &  \multicolumn{2}{c|}{\Comix} & \multicolumn{2}{c|}{\Chili} & \multicolumn{2}{c}{\Chili+NF} \\
       {\footnotesize(color} & $\Delta\sigma/\sigma$ & $\eta$ & $\Delta\sigma/\sigma$ & $\eta$ \\[-1mm]
       {\footnotesize\; sum)} & {\scriptsize 6M pts} & {\scriptsize 100 evts} & {\scriptsize 6M pts} &
       {\scriptsize 100 evts} & {\scriptsize 6M pts} & {\scriptsize 100 evts} \\\hline 
       $W^+$+1j & 0.4\textperthousand & $2\times 10^{-1}$ & 0.5\textperthousand & $2\times 10^{-1}$ & 0.2\textperthousand & $4\times10^{-1}$\\
       $W^+$+2j & 0.9\textperthousand & $2\times 10^{-2}$ & 0.7\textperthousand & $4\times 10^{-2}$ & 0.7\textperthousand & $5\times10^{-2}$\\
  \end{tabularx}\hfill
  \begin{tabularx}{0.4925\textwidth}{>{\raggedright\arraybackslash}X|
    >{\centering\arraybackslash}X|>{\centering\arraybackslash}X| 
    >{\centering\arraybackslash}X|>{\centering\arraybackslash}X| 
    >{\centering\arraybackslash}X|>{\centering\arraybackslash}X}
       Process &  \multicolumn{2}{c|}{\Comix} & \multicolumn{2}{c|}{\Chili} & \multicolumn{2}{c}{\Chili+NF} \\
       {\footnotesize(color} & $\Delta\sigma/\sigma$ & $\eta$ & $\Delta\sigma/\sigma$ & $\eta$ \\[-1mm]
       {\footnotesize\; sum)} & {\scriptsize 6M pts} & {\scriptsize 100 evts} & {\scriptsize 6M pts} &
       {\scriptsize 100 evts} & {\scriptsize 6M pts} & {\scriptsize 100 evts} \\\hline 
       $Z$+1j & 0.4\textperthousand & $3\times 10^{-1}$ & 0.4\textperthousand & $2\times 10^{-1}$ & 0.1\textperthousand & $5\times10^{-1}$\\
       $Z$+2j & 0.7\textperthousand & $4\times 10^{-2}$ & 0.7\textperthousand & $5\times 10^{-2}$ & 0.6\textperthousand & $6\times10^{-2}$\\
  \end{tabularx}\\\vskip 3mm
  \begin{tabularx}{0.4925\textwidth}{>{\raggedright\arraybackslash}X|
    >{\centering\arraybackslash}X|>{\centering\arraybackslash}X| 
    >{\centering\arraybackslash}X|>{\centering\arraybackslash}X| 
    >{\centering\arraybackslash}X|>{\centering\arraybackslash}X}
       Process &  \multicolumn{2}{c|}{\Comix} & \multicolumn{2}{c|}{\Chili} & \multicolumn{2}{c}{\Chili+NF} \\
       {\footnotesize(color} & $\Delta\sigma/\sigma$ & $\eta$ & $\Delta\sigma/\sigma$ & $\eta$ \\[-1mm]
       {\footnotesize\; sum)} & {\scriptsize 6M pts} & {\scriptsize 100 evts} & {\scriptsize 6M pts} &
       {\scriptsize 100 evts} & {\scriptsize 6M pts} & {\scriptsize 100 evts} \\\hline 
       $h$+1j & 0.2\textperthousand & $4\times 10^{-1}$ & 0.2\textperthousand & $5\times 10^{-1}$ & 0.05\textperthousand & $8\times10^{-1}$\\
       $h$+2j & 0.6\textperthousand & $6\times 10^{-2}$ & 0.3\textperthousand & $1\times 10^{-1}$ & 0.3\textperthousand & $2\times10^{-1}$\\
  \end{tabularx}\hfill
  \begin{tabularx}{0.4925\textwidth}{>{\raggedright\arraybackslash}X|
    >{\centering\arraybackslash}X|>{\centering\arraybackslash}X| 
    >{\centering\arraybackslash}X|>{\centering\arraybackslash}X| 
    >{\centering\arraybackslash}X|>{\centering\arraybackslash}X}
       Process &  \multicolumn{2}{c|}{\Comix} & \multicolumn{2}{c|}{\Chili} & \multicolumn{2}{c}{\Chili+NF} \\
       {\footnotesize(color} & $\Delta\sigma/\sigma$ & $\eta$ & $\Delta\sigma/\sigma$ & $\eta$ \\[-1mm]
       {\footnotesize\; sum)} & {\scriptsize 6M pts} & {\scriptsize 100 evts} & {\scriptsize 6M pts} &
       {\scriptsize 100 evts} & {\scriptsize 6M pts} & {\scriptsize 100 evts} \\\hline 
       $t\bar{t}$+0j & 0.2\textperthousand & $5\times 10^{-1}$ & 0.1\textperthousand & $6\times 10^{-1}$ & 0.05\textperthousand & $7\times10^{-1}$\\
       $t\bar{t}$+1j & 0.5\textperthousand & $1\times 10^{-1}$ & 0.2\textperthousand & $3\times 10^{-1}$ & 0.3\textperthousand & $2\times10^{-1}$\\
  \end{tabularx}\\\vskip 3mm
  \begin{tabularx}{0.4925\textwidth}{>{\raggedright\arraybackslash}X|
    >{\centering\arraybackslash}X|>{\centering\arraybackslash}X| 
    >{\centering\arraybackslash}X|>{\centering\arraybackslash}X| 
    >{\centering\arraybackslash}X|>{\centering\arraybackslash}X}
       Process &  \multicolumn{2}{c|}{\Comix} & \multicolumn{2}{c|}{\Chili} & \multicolumn{2}{c}{\Chili+NF} \\
       {\footnotesize(color} & $\Delta\sigma/\sigma$ & $\eta$ & $\Delta\sigma/\sigma$ & $\eta$ \\[-1mm]
       {\footnotesize\; sum)} & {\scriptsize 6M pts} & {\scriptsize 100 evts} & {\scriptsize 6M pts} &
       {\scriptsize 100 evts} & {\scriptsize 6M pts} & {\scriptsize 100 evts} \\\hline 
       $\gamma$+1j & 0.3\textperthousand & $4\times 10^{-1}$ & 0.7\textperthousand & $2\times 10^{-1}$ & 0.1\textperthousand & $5\times10^{-1}$\\
       $\gamma$+2j & 1.0\textperthousand & $1\times 10^{-2}$ & 1.9\textperthousand & $5\times 10^{-3}$ & 1.4\textperthousand & $9\times 10^{-3}$\\
  \end{tabularx}\hfill
  \begin{tabularx}{0.4925\textwidth}{>{\raggedright\arraybackslash}X|
    >{\centering\arraybackslash}X|>{\centering\arraybackslash}X| 
    >{\centering\arraybackslash}X|>{\centering\arraybackslash}X| 
    >{\centering\arraybackslash}X|>{\centering\arraybackslash}X}
       Process &  \multicolumn{2}{c|}{\Comix} & \multicolumn{2}{c|}{\Chili} & \multicolumn{2}{c}{\Chili+NF} \\
       {\footnotesize(color} & $\Delta\sigma/\sigma$ & $\eta$ & $\Delta\sigma/\sigma$ & $\eta$ \\[-1mm]
       {\footnotesize\; sum)} & {\scriptsize 6M pts} & {\scriptsize 100 evts} & {\scriptsize 6M pts} &
       {\scriptsize 100 evts} & {\scriptsize 6M pts} & {\scriptsize 100 evts} \\\hline 
       2jets & 0.4\textperthousand & $2\times 10^{-1}$ & 0.2\textperthousand & $4\times 10^{-1}$ & 0.08\textperthousand & $6\times10^{-1}$\\
       3jets & 0.8\textperthousand & $2\times 10^{-2}$ & 0.6\textperthousand & $6\times 10^{-2}$ & 0.7\textperthousand & $3\times10^{-2}$\\
  \end{tabularx}\\
  \caption{Relative Monte-Carlo uncertainties, $\Delta\sigma/\sigma$, and unweighting efficiencies,
    $\eta$, in color-summed leading-order calculations. The center-of-mass energy is $\sqrt{s}=14$~TeV,
    jets are defined using the anti-$k_T$ algorithm with $p_{\perp,j}=30$~GeV and $|y_j|\le 6$.
    For details see the main text.}
  \label{tab:comparison_madnis}
\end{table}
\begin{table}[t]
  \begin{tabularx}{0.4925\textwidth}{>{\raggedright\arraybackslash}X|
    >{\centering\arraybackslash}X|>{\centering\arraybackslash}X| 
    >{\centering\arraybackslash}X|>{\centering\arraybackslash}X| 
    >{\centering\arraybackslash}X|>{\centering\arraybackslash}X}
       Process &  \multicolumn{2}{c|}{\Comix} & \multicolumn{2}{c|}{\Chili} & \multicolumn{2}{c}{\Chili+NF} \\
       {\footnotesize(color} & Opt & Gen & Opt & Gen & Opt & Gen \\[-1mm]
       {\footnotesize\; sum)} & {\scriptsize 0.8M pts} & {\scriptsize 6M pts} & {\scriptsize 0.8M pts} &
       {\scriptsize 6M pts} & {\scriptsize 1.2M pts} &
       {\scriptsize 6M pts} \\\hline 
       $W^+$+1j & 2m & 10m & 1m & 8m & 5m & 8m \\
       $W^+$+2j & 14m & 1.9h & 13m & 1.7h & 29m & 1.3h \\
  \end{tabularx}\hfill
  \begin{tabularx}{0.4925\textwidth}{>{\raggedright\arraybackslash}X|
    >{\centering\arraybackslash}X|>{\centering\arraybackslash}X| 
    >{\centering\arraybackslash}X|>{\centering\arraybackslash}X| 
    >{\centering\arraybackslash}X|>{\centering\arraybackslash}X}
       Process &  \multicolumn{2}{c|}{\Comix} & \multicolumn{2}{c|}{\Chili} & \multicolumn{2}{c}{\Chili+NF} \\
       {\footnotesize(color} & Opt & Gen & Opt & Gen & Opt & Gen \\[-1mm]
       {\footnotesize\; sum)} & {\scriptsize 0.8M pts} & {\scriptsize 6M pts} & {\scriptsize 0.8M pts} &
       {\scriptsize 6M pts} & {\scriptsize 1.2M pts} &
       {\scriptsize 6M pts} \\\hline 
       $Z$+1j & 2m & 19m & 2m & 14m & 7m & 15m \\
       $Z$+2j & 30m & 3.9h & 20m & 3.3h & 58m & 2.9h \\
  \end{tabularx}\\\vskip 3mm
  \begin{tabularx}{0.4925\textwidth}{>{\raggedright\arraybackslash}X|
    >{\centering\arraybackslash}X|>{\centering\arraybackslash}X| 
    >{\centering\arraybackslash}X|>{\centering\arraybackslash}X| 
    >{\centering\arraybackslash}X|>{\centering\arraybackslash}X}
       Process &  \multicolumn{2}{c|}{\Comix} & \multicolumn{2}{c|}{\Chili} & \multicolumn{2}{c}{\Chili+NF} \\
       {\footnotesize(color} & Opt & Gen & Opt & Gen & Opt & Gen \\[-1mm]
       {\footnotesize\; sum)} & {\scriptsize 0.8M pts} & {\scriptsize 6M pts} & {\scriptsize 0.8M pts} &
       {\scriptsize 6M pts} & {\scriptsize 1.2M pts} &
       {\scriptsize 6M pts} \\\hline 
       $h$+1j & 1m & 10m & 1m & 7m & 4m & 8m \\
       $h$+2j & 8m & 1.1h & 6m & 52m & 18m & 46m \\
  \end{tabularx}\hfill
  \begin{tabularx}{0.4925\textwidth}{>{\raggedright\arraybackslash}X|
    >{\centering\arraybackslash}X|>{\centering\arraybackslash}X| 
    >{\centering\arraybackslash}X|>{\centering\arraybackslash}X| 
    >{\centering\arraybackslash}X|>{\centering\arraybackslash}X}
       Process &  \multicolumn{2}{c|}{\Comix} & \multicolumn{2}{c|}{\Chili} & \multicolumn{2}{c}{\Chili+NF} \\
       {\footnotesize(color} & Opt & Gen & Opt & Gen & Opt & Gen \\[-1mm]
       {\footnotesize\; sum)} & {\scriptsize 0.8M pts} & {\scriptsize 6M pts} & {\scriptsize 0.8M pts} &
       {\scriptsize 6M pts} & {\scriptsize 1.2M pts} &
       {\scriptsize 6M pts} \\\hline 
       $t\bar{t}$+0j & 1m & 9m & 1m & 6m & 4m & 7m \\
       $t\bar{t}$+1j & 6m & 54m & 6m & 42m & 17m & 40m \\
  \end{tabularx}\\\vskip 3mm
  \begin{tabularx}{0.4925\textwidth}{>{\raggedright\arraybackslash}X|
    >{\centering\arraybackslash}X|>{\centering\arraybackslash}X| 
    >{\centering\arraybackslash}X|>{\centering\arraybackslash}X| 
    >{\centering\arraybackslash}X|>{\centering\arraybackslash}X}
       Process &  \multicolumn{2}{c|}{\Comix} & \multicolumn{2}{c|}{\Chili} & \multicolumn{2}{c}{\Chili+NF} \\
       {\footnotesize(color} & Opt & Gen & Opt & Gen & Opt & Gen \\[-1mm]
       {\footnotesize\; sum)} & {\scriptsize 0.8M pts} & {\scriptsize 6M pts} & {\scriptsize 0.8M pts} &
       {\scriptsize 6M pts} & {\scriptsize 1.2M pts} &
       {\scriptsize 6M pts} \\\hline 
       $\gamma$+1j & 2m & 15m & 1m & 11m & 6m & 13m \\
       $\gamma$+2j & 22m & 2.9h & 19m & 2.2h & 38m & 2.0h \\
  \end{tabularx}\hfill
  \begin{tabularx}{0.4925\textwidth}{>{\raggedright\arraybackslash}X|
    >{\centering\arraybackslash}X|>{\centering\arraybackslash}X| 
    >{\centering\arraybackslash}X|>{\centering\arraybackslash}X| 
    >{\centering\arraybackslash}X|>{\centering\arraybackslash}X}
       Process &  \multicolumn{2}{c|}{\Comix} & \multicolumn{2}{c|}{\Chili} & \multicolumn{2}{c}{\Chili+NF} \\
       {\footnotesize(color} & Opt & Gen & Opt & Gen & Opt & Gen \\[-1mm]
       {\footnotesize\; sum)} & {\scriptsize 0.8M pts} & {\scriptsize 6M pts} & {\scriptsize 0.8M pts} &
       {\scriptsize 6M pts} & {\scriptsize 1.2M pts} &
       {\scriptsize 6M pts} \\\hline 
       2jets & 6m & 47m & 5m & 37m & 14m & 34m \\
       3jets & 27m & 3.6h & 24m & 3.0h & 45m & 2.6h \\
  \end{tabularx}\\
  \caption{Time for optimization and event generation in leading-order calculations. The center-of-mass energy is $\sqrt{s}=14$~TeV, jets are defined
    using the anti-$k_T$ algorithm with $p_{\perp,j}=30$~GeV and $|y_j|\le 6$. The optimization step consists of 0.8 or 1.2 million non-zero events and the generation
    consists of 6 million non-zero, weighted events.
    All codes are generated using dual-socket eight-core Intel E5-2650v2 ``Ivy Bridge'' (2.6 GHz) CPUs.
    The \Chili+NF results are obtained using a single threaded version of \Sherpa and 16 cores for the optimization of the NF parameters in Tensorflow.
    To be consistent with \Chili+NF, the results for \Sherpa and \Chili are total runtime, summing over all MPI ranks.
    }
  \label{tab:comparison_madnis_time}
\end{table}
In this section we present first numerical results obtained with our new integrator, \Chili.
We have interfaced the new framework with the general-purpose event generator
\Sherpa~\cite{Gleisberg:2003xi,Gleisberg:2008ta,Sherpa:2019gpd}, which is used 
to compute the partonic matrix elements and the parton luminosity with the help of
\Comix~\cite{Gleisberg:2008fv} and Amegic~\cite{Krauss:2001iv}. To allow performance tests 
from low to high particle multiplicity, we use Comix' default method of sampling of the 
QCD color space~\cite{Duhr:2006iq,Gleisberg:2008fv}, unless explicitly stated otherwise.
This affects the convergence rate, and we note that better MC uncertainties could in principle 
be obtained for color-summed computations, but at the cost of much larger computing time at 
high multiplicity. The performance comparison between \Sherpa and \Chili would, 
however, be unaffected. We use the NNPDF~3.0 PDF set~\cite{NNPDF:2014otw} 
at NNLO precision, and the corresponding settings of the strong coupling, i.e.\ 
$\alpha_s(m_z)=0.118$ and running to 3-loop order.
Light quarks, charm and bottom quarks are assumed to be massless, and we set $m_t=173.21$.
The electroweak parameters are determined in the complex mass scheme using the
inputs $\alpha(m_Z)=1/128.8$, $m_W=80.385$, $m_Z=91.1876$, $m_h=125$ and 
$\Gamma_W=2.085$, $\Gamma_Z=2.4952$.
We assume incoming proton beams at a hadronic center-of-mass energy of $\sqrt{s}=14$~TeV.
To implement basic phase-space cuts, we reconstruct jets using the anti-$k_T$ jet
algorithm~\cite{Cacciari:2008gp} with $R=0.4$ in the implementation of FastJet~\cite{Cacciari:2011ma}
and require $p_{\perp,j}\ge30$~GeV and $|y_j|\le6$. 
Photons are isolated from QCD activity based on Ref.~\cite{Frixione:1998jh}
with $\delta_0$=0.4, $n$=2 and $\epsilon_\gamma$=2.5\% and are required 
to have $p_{\perp,\gamma}\ge30$~GeV.
All results presented in this section are obtained with a scalable version of our
new integrator using parallel execution on CPUs with the help of MPI.

Table~\ref{tab:comparison_democratic} shows a comparison between MC uncertainties 
and event generation efficiencies in leading-order calculations, obtained with the
recursive phase-space generator in \Comix and with \Chili. A brief description of the
recursive phase-space integrator implemented in Comix is given 
in App.~\ref{sec:recursive_ps}. To improve the convergence 
of the integrals we use the Vegas~\cite{Lepage:1977sw} algorithm, which is implemented
independently in both \Sherpa and \Chili. The MC uncertainties are given after optimizing
the adaptive integrator with 1.2 million non-zero phase-space points and evaluation 
of the integral with 6 million non-zero phase-space points. We employ the definition
of event generation efficiency in Ref.~\cite{Gao:2020zvv}, and we evaluate it using
100 replicas of datasets leading to 100 unweighted events each.
For more details on our definition of event generation efficiency see App.~\ref{App:ps_efficiency}.
We test the production 
of $W^+$ and $Z$ bosons with leptonic decay, on-shell Higgs boson production, 
top-quark pair production, direct photon production and pure QCD jet production.
These processes are omnipresent in background simulations at the Large Hadron Collider (LHC),
and are typically associated with additional light jet activity due to the large phase space.
Accordingly, we test the basic process with up to four additional light jets, 
where all additional radiated jets are assumed to be purely from QCD interactions 
and do not include additional electroweak bosons. In single boson production we 
do not include the trivial process without any light jets. We observe that the 
performance of our new integrator is well comparable to that of the recursive 
phase-space generator in \Sherpa, especially for less than 5 additional jets 
with the exception of $\gamma+$jets. In many cases it shows slightly higher
unweighting efficiencies. This is both
encouraging and somewhat surprising, given the relative simplicity of our new
approach, which does not make use of repeated t-channel factorization.
Due to the uniform jet cuts, we even obtain similar performance
when using the minimal number of s-channel parametrizations,
where the minimal number is 1 for $W$ and $Z$ processes and 0 otherwise.
This setup is labeled as \Chili (basic) in Tab.~\ref{tab:comparison_democratic}.
The results suggest that a single phase-space parametrization may in many cases be
sufficient to compute cross sections and generate events at high precision,
which is advantageous in terms of computing time and helps to scale the computation
to higher multiplicity processes. Moreover, it circumvents the problems related 
to multi-channel integration discussed in~\cite{Gao:2020zvv,Heimel:2022wyj} 
when combining our integrator with neural network based adaptive random number 
mapping techniques. We note that this configuration is also used by MCFM~\cite{Campbell:2003hd}.

Table~\ref{tab:comparison_boosted} shows a similar comparison as in
Tab.~\ref{tab:comparison_democratic}, but in addition we apply a cut
on the leading jet, requiring $p_{\perp,j1}>300$~GeV. This configuration
tests the regime where the hard system receives a large boost, and there is
usually a strong hierarchy between the jet transverse momenta. In these
scenarios we expect the complete \Chili integrator to outperform the basic 
configuration with a t-channel only, which is confirmed by the comparison
in Tab.~\ref{tab:comparison_boosted}.
The only exception to this is the $\gamma+5j$ process,
which may be a result of the poor integration accuracy from \Chili for this process.
The lower right sub-table shows a 
configuration where we do not apply the additional transverse momentum cut,
but instead use a large di-jet invariant mass cut, typical for VBF searches
and measurements, $m_{j1,j2}\ge600$~GeV.
Here we see that \Chili and \Chili (basic) are roughly comparable
and perform better than the default \Sherpa integrator, with the 
exception of $h+2j$.

Table~\ref{tab:comparison_nlo} shows a comparison of MC uncertainties
and cut efficiencies for various next-to-leading order QCD computations. 
We use the Catani-Seymour dipole subtraction method~\cite{Catani:1996vz},
where the value of an arbitrary infrared-safe observable, $O$, can be computed
with the help of the Born differential cross section, $B$, the UV renormalized
virtual  corrections, $V$, the collinear mass factorization counterterms, $C$,
and a set of differential and integrated infrared subtraction counterterms,
$D_i$ and $I_i=\int{\rm d}\Phi_{+1}D_i$, where ${\rm d}\Phi_{+1}$ is the 
differential one-emission phase space associated with the production of an
additional parton~\cite{Catani:1996vz}:
\begin{equation}\label{eq:nlo_master}
  \begin{split}
    \langle O\rangle=&\;\int{\rm d}\Phi_n\left[
    B(\Phi_n)+V(\Phi_n)+C(\Phi_n)+\sum_iI_i(\Phi_n)\right]O(\Phi_n)\\
    &\qquad+\int{\rm d}\Phi_{n+1}\left[R(\Phi_{n+1})O(\Phi_{n+1})
    -\sum_iD_i(\Phi_{n+1})O(\Phi_{n,i})\right]\;.
  \end{split}
\end{equation}
We note that in the second integral, the value of the observable, $O$,
is computed based on the real-emission phase-space point in the first term,
and based on the projected Born-like phase-space points in the dipole
subtration terms. Each dipole term has its own, specific projection.
The fact that the cancelation of infrared enhancements in the second
integral occurs non-locally in phase space makes the evaluation particularly
cumbersome with Monte-Carlo methods. While the associated integral is finite
for any infrared safe observable, it typically has large Monte-Carlo uncertainties
due to imperfect cancelations of positive and negative contributions.

We assign the shorthand B-like for first line in Eq.~\eqref{eq:nlo_master},
and the shorthand R-like for the second line. Both calculations exhibit
different structures than at leading order in QCD, cf.~\cite{Ellis:2009zw}. 
The real-emission integrals in particular test the efficiency of the dipole mapping
described in Sec.~\ref{sec:dipole_mapping}, which is designed to match
the structure of the differential infrared counterterms, $D_i$.
It can be seen that our new algorithm has a much better cut efficiency than
the recursive phase-space generator in \Sherpa, which is again advantageous
in terms of overall computing time. The cut efficiency, $\varepsilon_{\rm cut}$
is defined as the ratio between the number of Monte-Carlo points that pass the
phase-space cuts, and the total number of points. The MC uncertainty for a given number 
of phase-space points is reduced at low jet multiplicity, and generally comparable
to the recursive phase-space generator. Given the simplicity of the \Chili approach,
this is a very encouraging result for the development of NLO simulations on
modern computing architectures. If a speedup of the matrix-element calculation
is obtained, for example through analytic expressions~\cite{Campbell:2021vlt},
accelerated numerical evaluation~\cite{Bothmann:2021nch,Valassi:2021ljk,
  Valassi:2022dkc,Bothmann:2022itv} or the usage of surrogate 
methods~\cite{Danziger:2021eeg,Janssen:2023ahv}, then the linear scaling
with the number of outgoing particles of the basic \Chili generator at leading order,
and the polynomial scaling with the number of outgoing particles of the 
dipole-based generator~\footnote{The number of dipole subtraction terms in the
Catani-Seymour method scales at most as $n^3$ with the number $n$ of external partons.},
will become an important feature.

Table~\ref{tab:comparison_madnis} shows a comparison of the Vegas-based \Chili integrator
and the neural-network assisted integrator for color summed matrix elements. 
We use the single channel configuration of \MadNIS~\cite{Heimel:2022wyj} (which is consistent with iFlow~\cite{Gao:2020vdv})
in combination with the basic \Chili integrator, while the Vegas-based version
of \Chili includes all possible $s$-channel mappings. The network is setup with 6 
rational quadratic spline coupling layers~\cite{durkan2019neural} with random permutations,
each consisting of a neural network with 2 layers with 16 nodes each using a leaky
ReLU activation function. 
In general, a coupling layer invertibly maps an input vector onto another one.
For an $n$-dimensional input vector $x$, let $A$ and $B$ denote two
disjoint sets of $\{1,\dots,d\}$. Then the coupling layer mapping is defined via
\begin{align}
  \begin{split}
    x^A &\mapsto y^A \coloneqq x^A\, , \\
    x^B &\mapsto y^B \coloneqq C\left( x^B; m(x^A)\right)\, ,    \label{eq:trafo_general}
  \end{split}
\end{align}
where $m$ is any function and $C$ is a separable, invertible function
on $\mathbb{R}^{|B|} \times m(x^A)$. Here $|B|$ denotes the cardinality of $B$, i.e.\ the
number of dimensions in the set $B$, and separability means that the mapping is applied
element-wise as
\begin{equation}
  C(x^B; m(x^A)) = \left( C_1\left(x_1^B; m(x^A)\right), \ldots,
    C_{|B|}\left(x_{|B|}^B; m(x^A)\right) \right)^T\,.
\end{equation}
The splitting of the input vector into two disjoint sets, where one set is being mapped
and the other set is used to parameterize the mapping combined with the requirements on the function $C$, allows for a simple computation of
the Jacobian determinant and crucially does not require
the inversion of $m$. This allows the usage of neural networks as functions $m$. For more details see e.g.\ Refs.~\cite{Gao:2020vdv,Bothmann:2020ywa}

The network is trained using 20
epochs of training with 100 batches of 1000 events per epoch with the variance as the loss term as in Ref.~\cite{Heimel:2022wyj}. The learning rate starts at 
0.001 and decays each epoch by $l_0 / ( 1 + l_d s / d_s)$, where $l_0$ is the initial 
learning rate, $l_d = 0.01$ is the decay rate, $s$ is the number of steps, and $d_s = 100$ is the 
number of steps before applying the decay.
Optimizing these parameters to achieve peak performance is beyond the scope of this project
and can be done in a similar fashion as in Ref.~\cite{Gao:2020zvv}.

The timings for \Sherpa, \Chili, and \Chili+NF are given in Tab.~\ref{tab:comparison_madnis_time}.
Here we find that the optimization of the normalizing flows is significantly slower than the Vegas optimization, even when including all of the possible phase space channels.
However, the generation of 6 million weighted events is approximately the same for the lowest multiplicity processes, but the normalizing flow approach does better after adding one additional jet.
This is a combination of having a better cut efficiency and significantly fewer channels.
We leave a more detailed investigation on the timing benefits to unweighting efficiency benefits at high multiplicities to a future work.

\begin{figure}[p]
    \centering
    \includegraphics[width=0.48\textwidth]{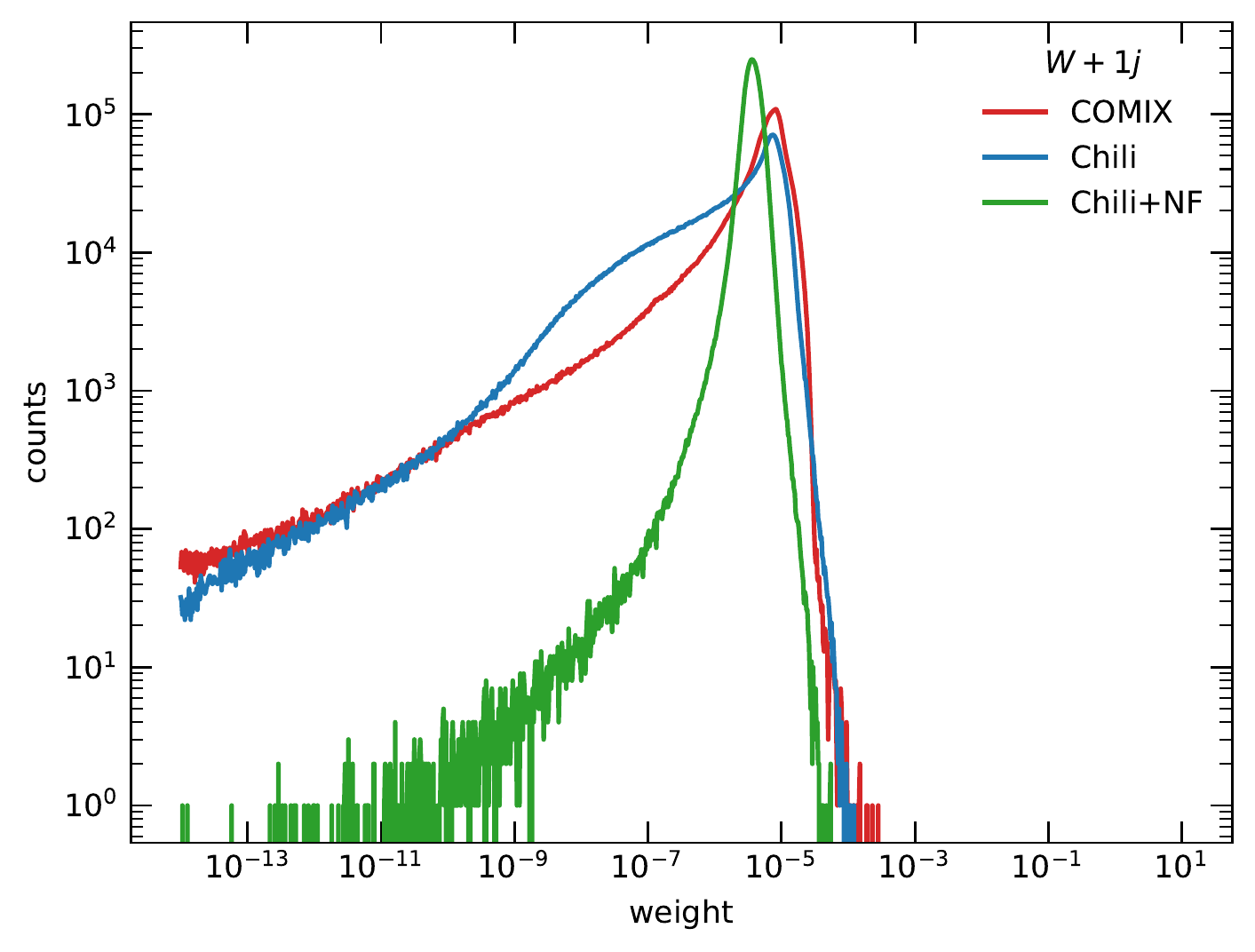}
    \hfill
    \includegraphics[width=0.48\textwidth]{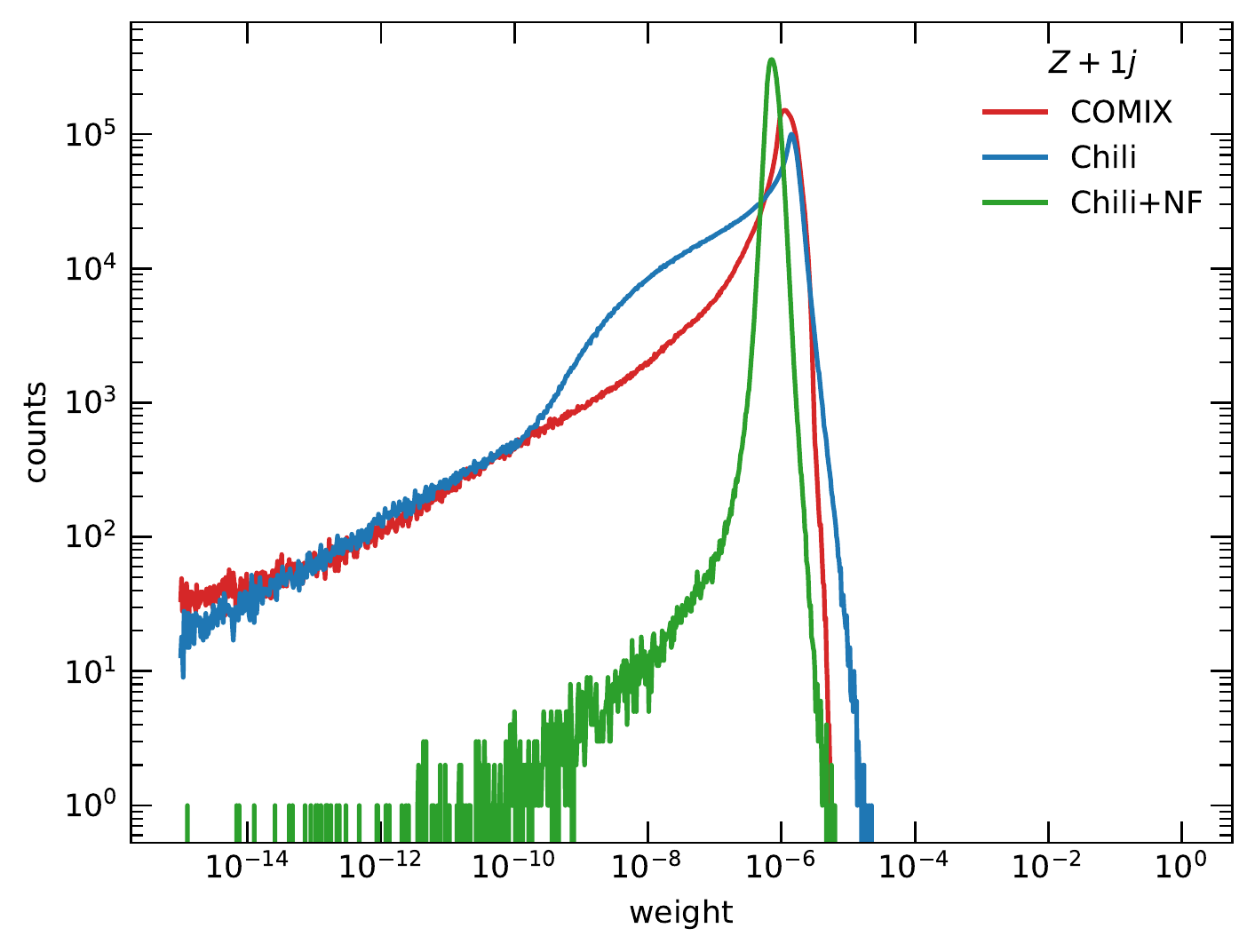} \\
    \includegraphics[width=0.48\textwidth]{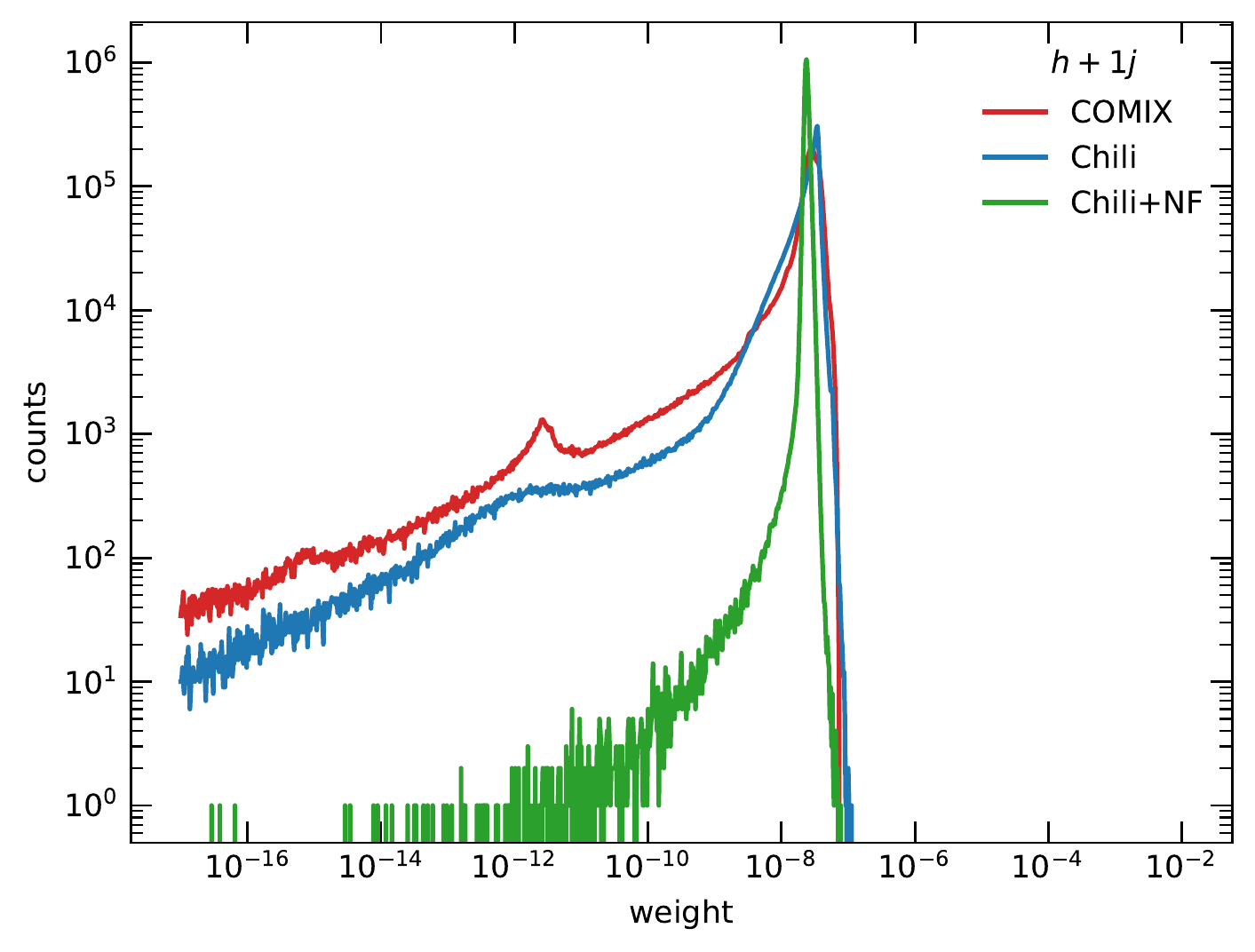}
    \hfill
    \includegraphics[width=0.48\textwidth]{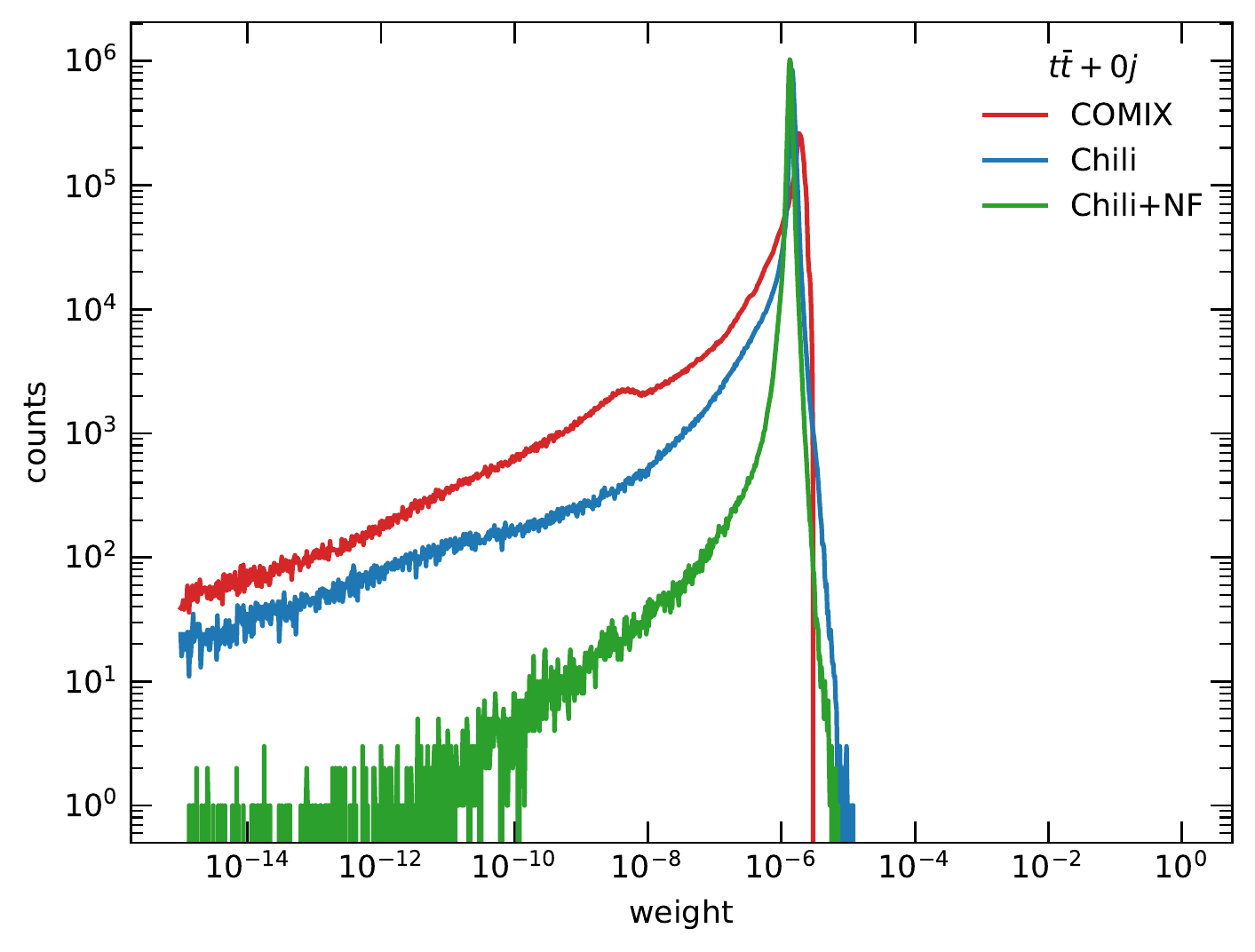} \\
    \includegraphics[width=0.48\textwidth]{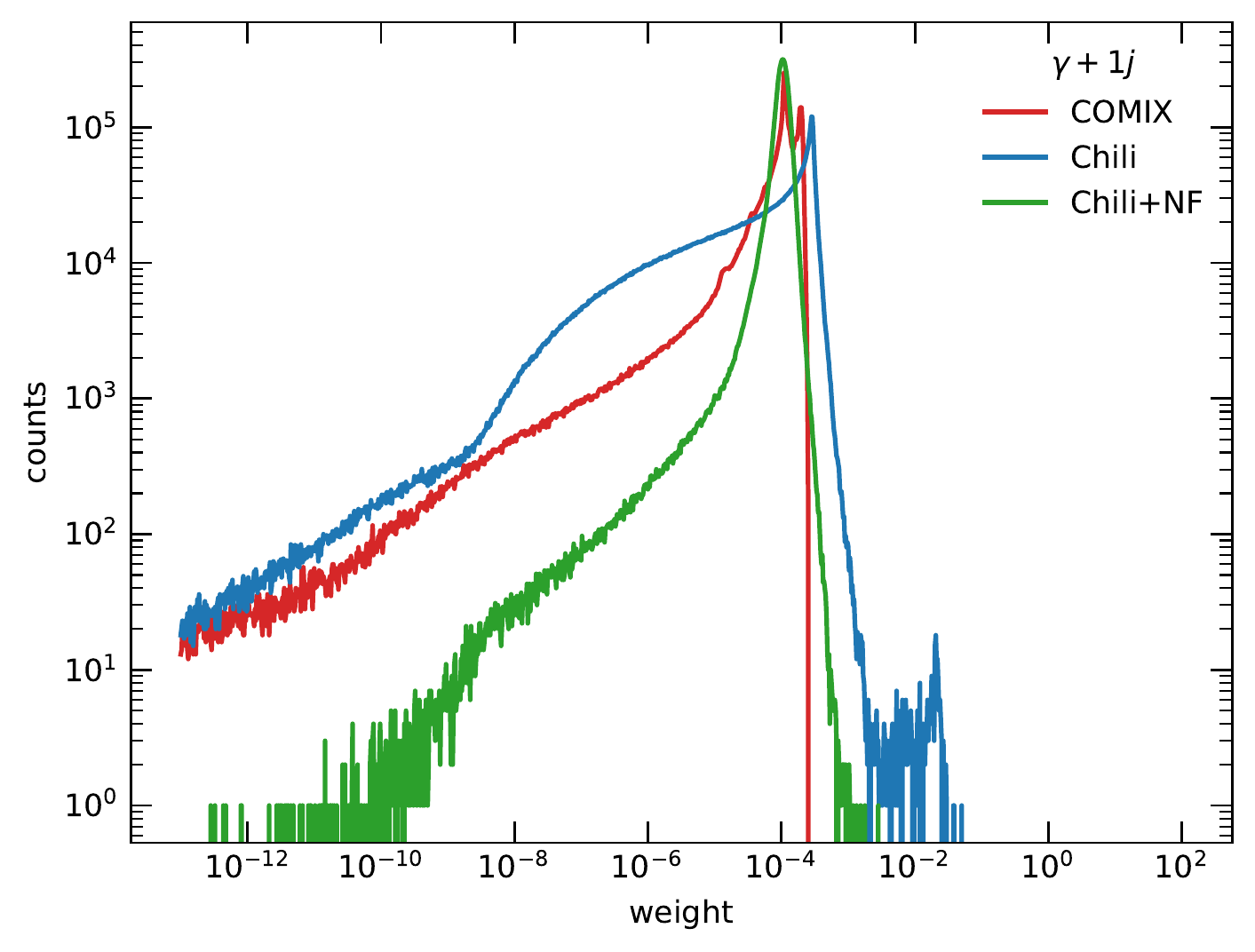}
    \hfill
    \includegraphics[width=0.48\textwidth]{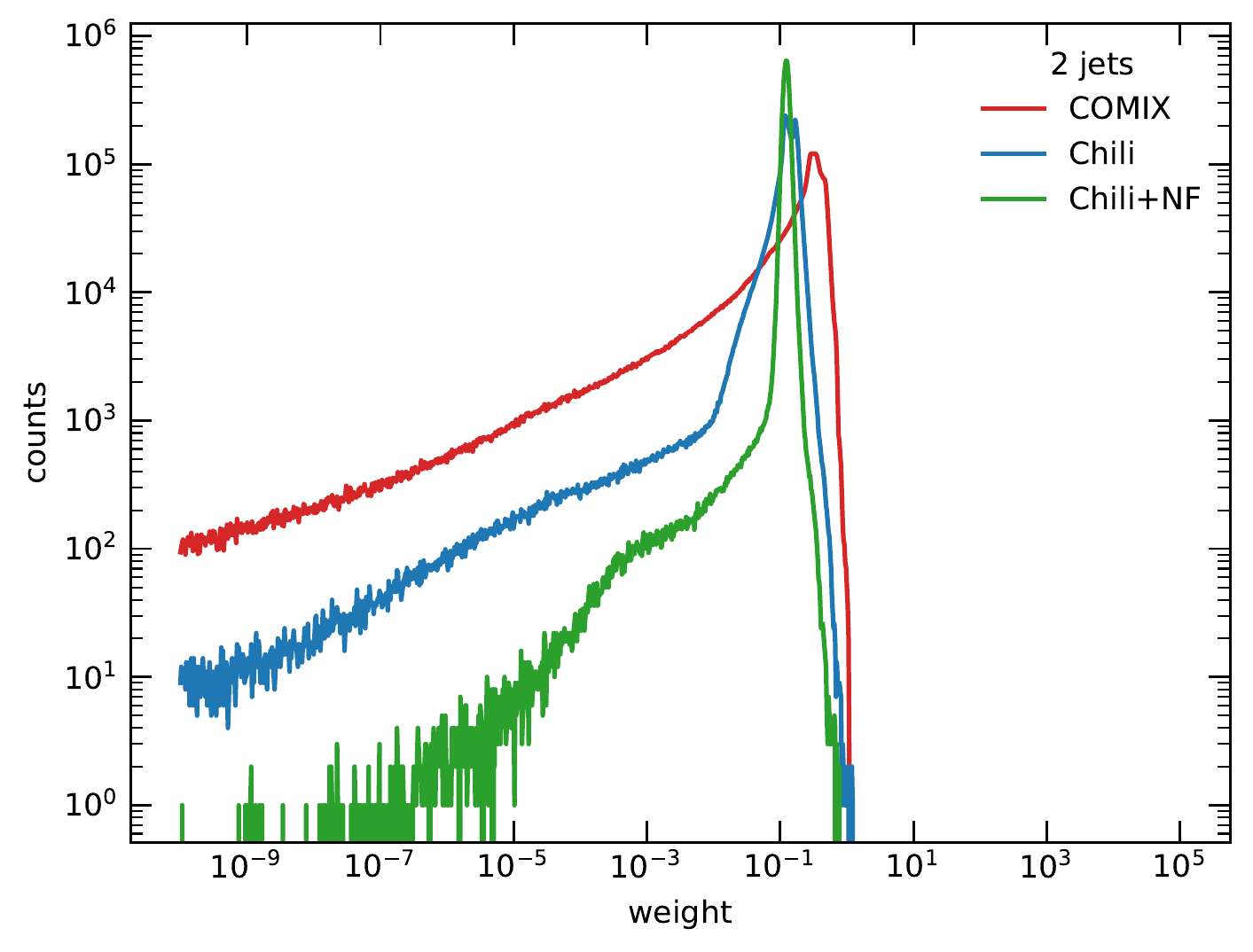}
    \caption{Weight distribution for the lowest multiplicity processes found in Tab.~\ref{tab:comparison_madnis}. Each curve contains 6 million events. The \Comix integrator is shown in red, the \Chili with Vegas is shown in blue, and \Chili with normalizing flows is shown in green. The results for $W+1j$ is in the upper right, $Z+1j$ in the upper left, the middle row consists of $h+1j$ and $t\bar{t}+0j$, and the bottom row has $\gamma+1j$ and dijets respectively.}
    \label{fig:comparison_madnis1}
\end{figure}
\begin{figure}[p]
    \centering
    \includegraphics[width=0.48\textwidth]{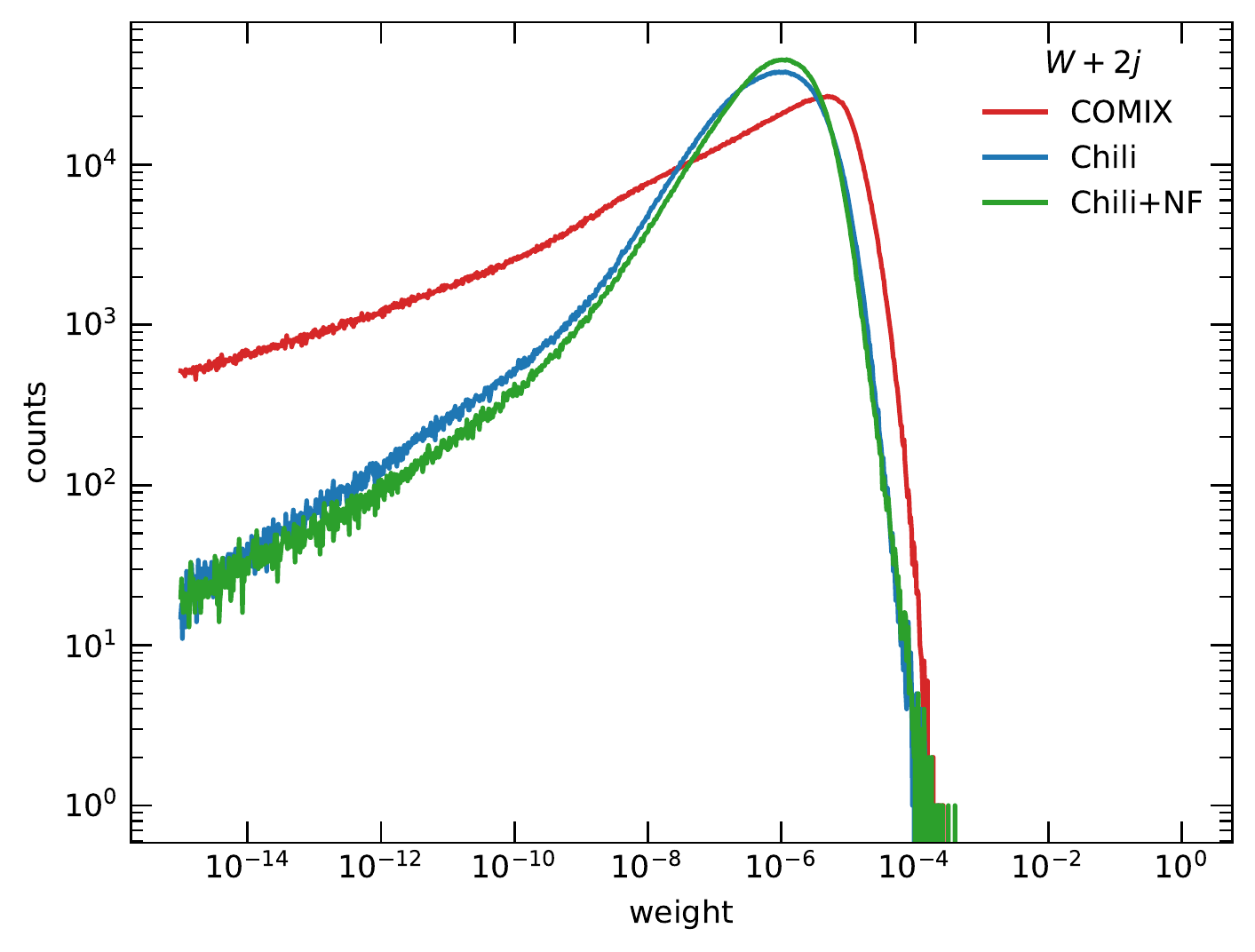}
    \hfill
    \includegraphics[width=0.48\textwidth]{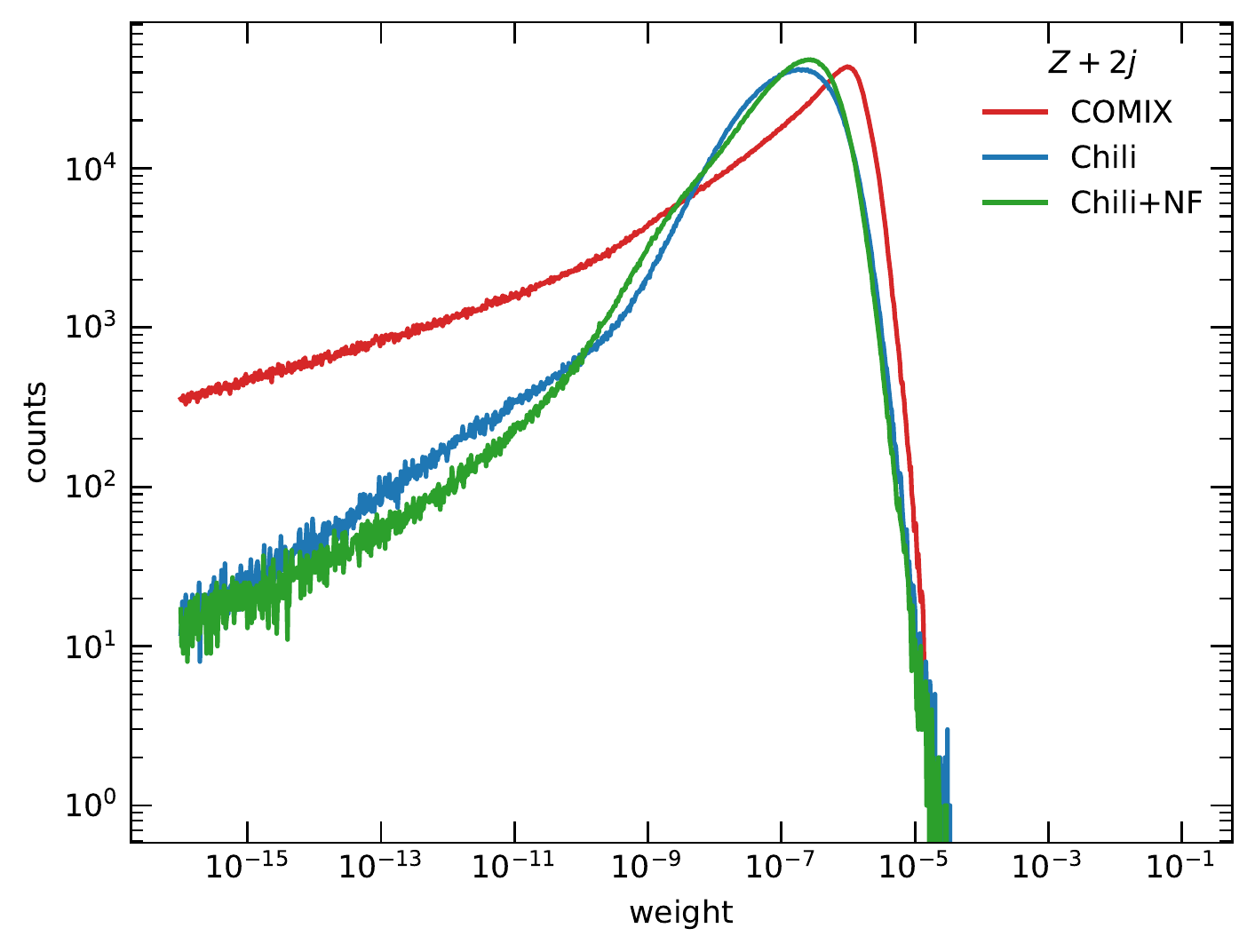} \\
    \includegraphics[width=0.48\textwidth]{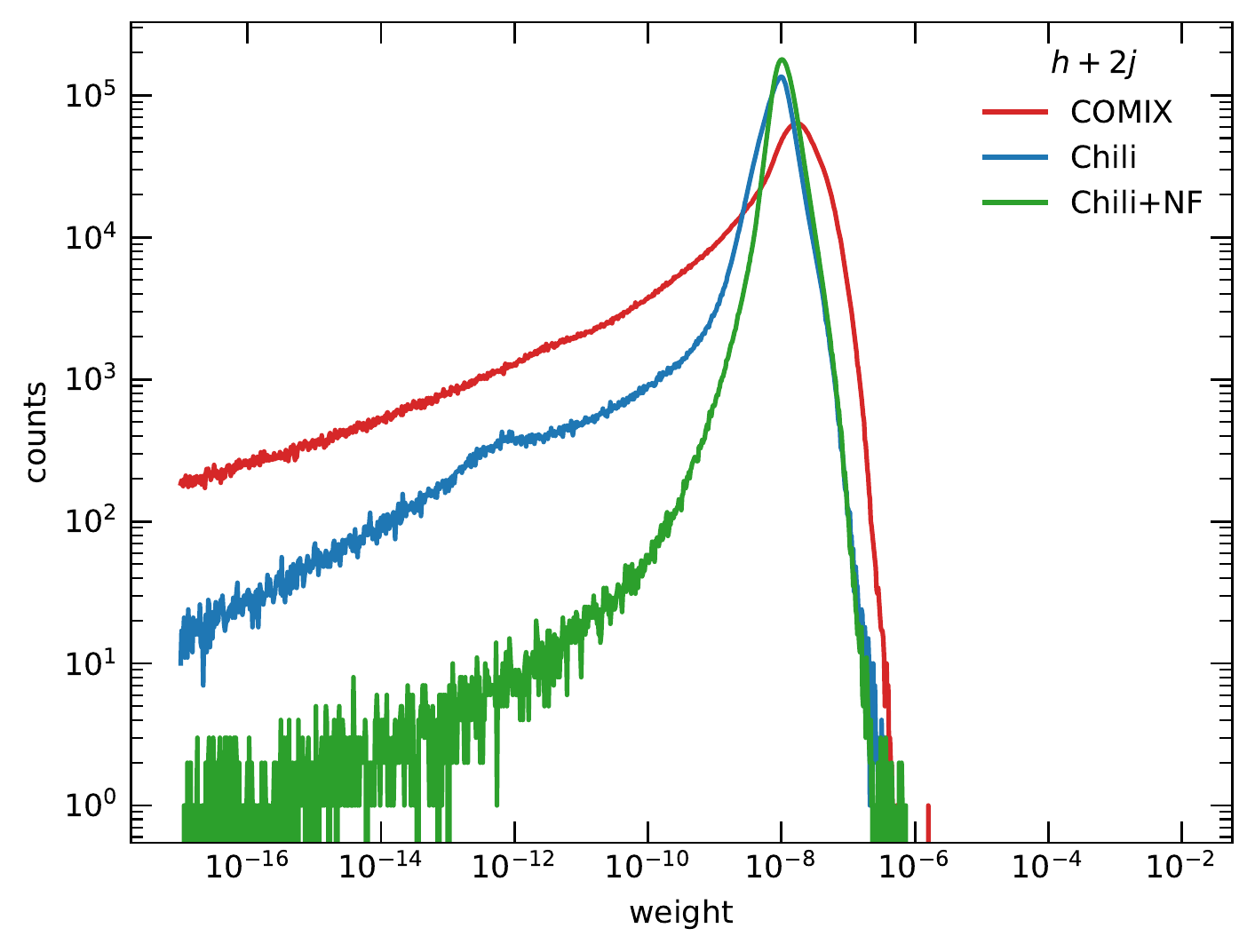}
    \hfill
    \includegraphics[width=0.48\textwidth]{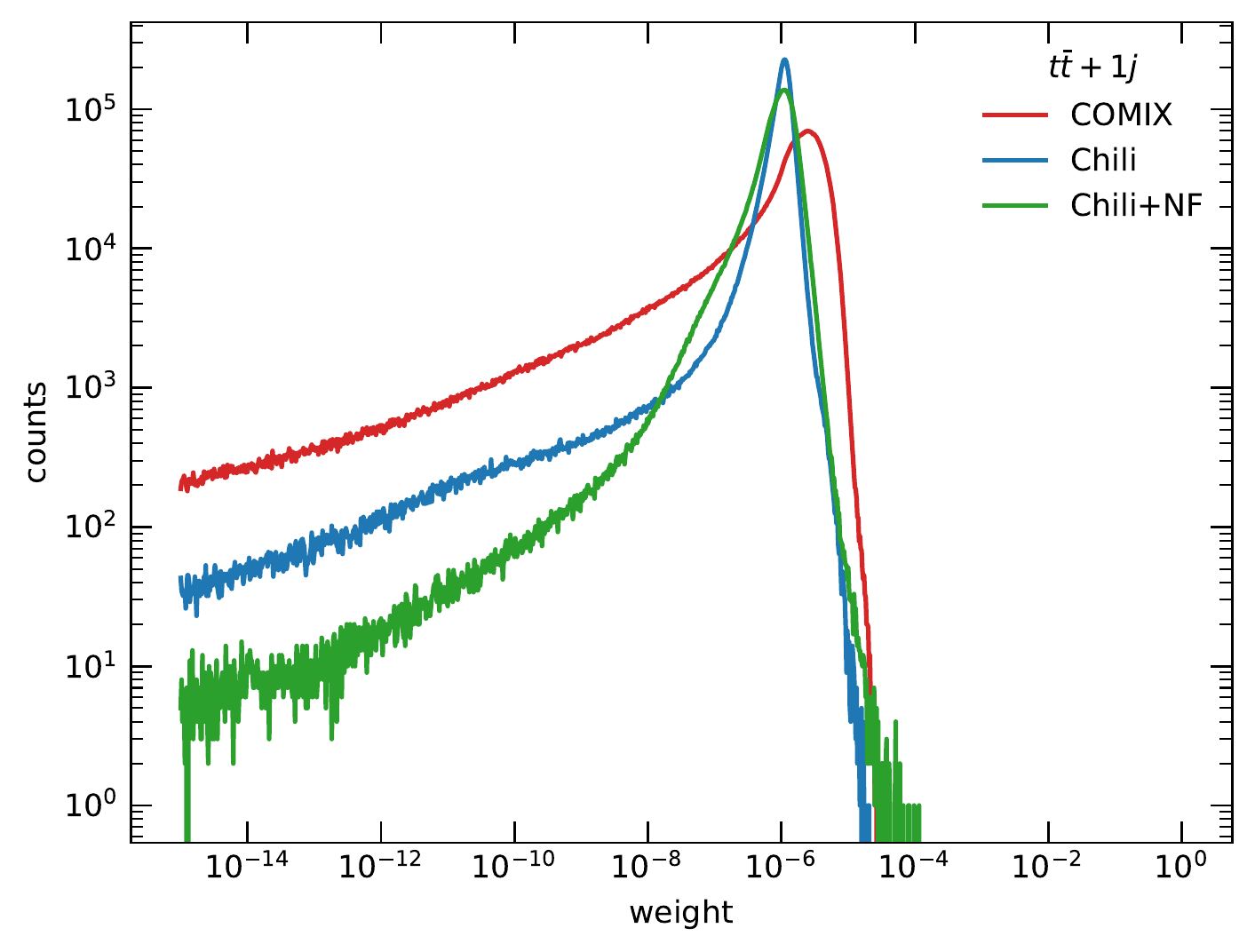} \\
    \includegraphics[width=0.48\textwidth]{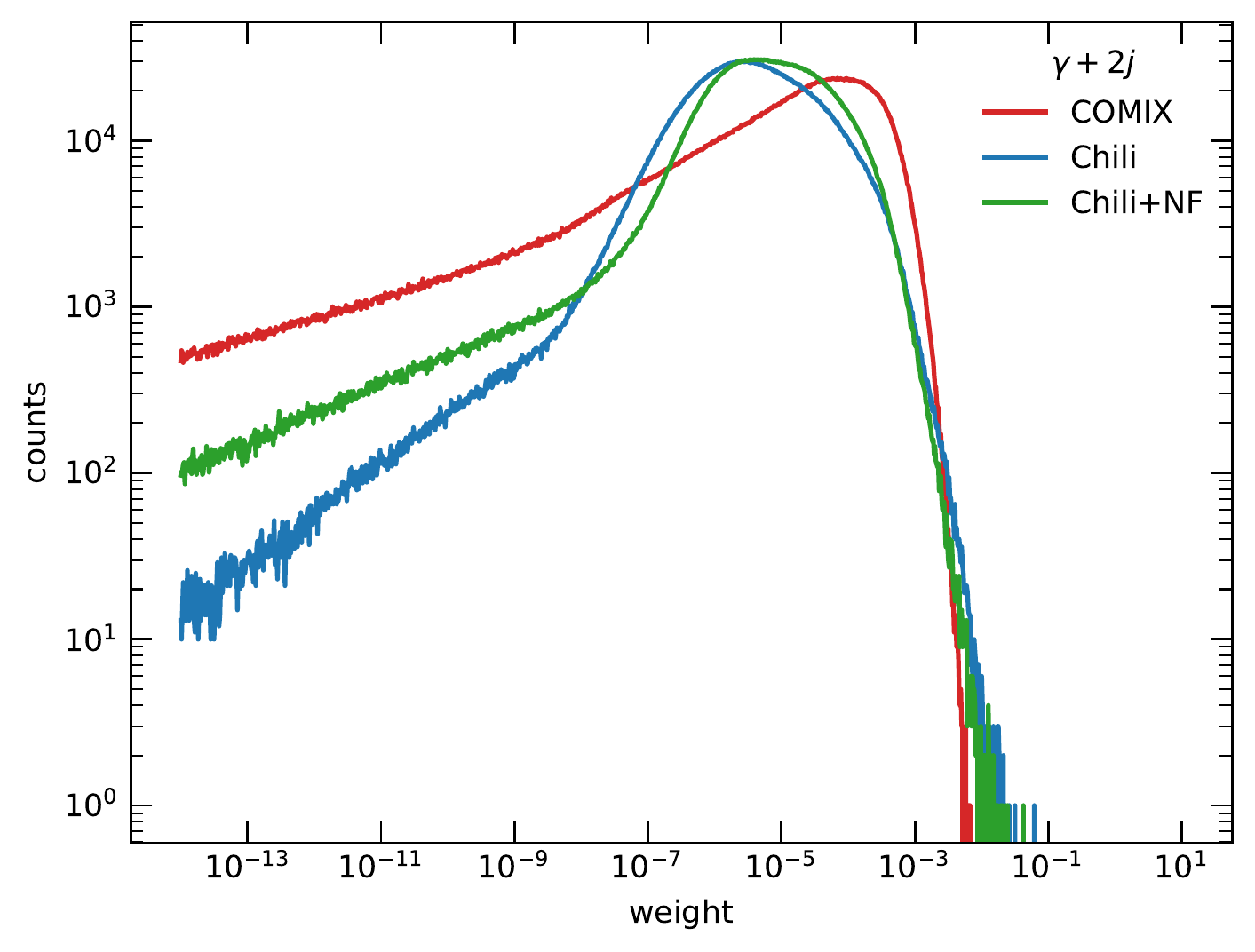}
    \hfill
    \includegraphics[width=0.48\textwidth]{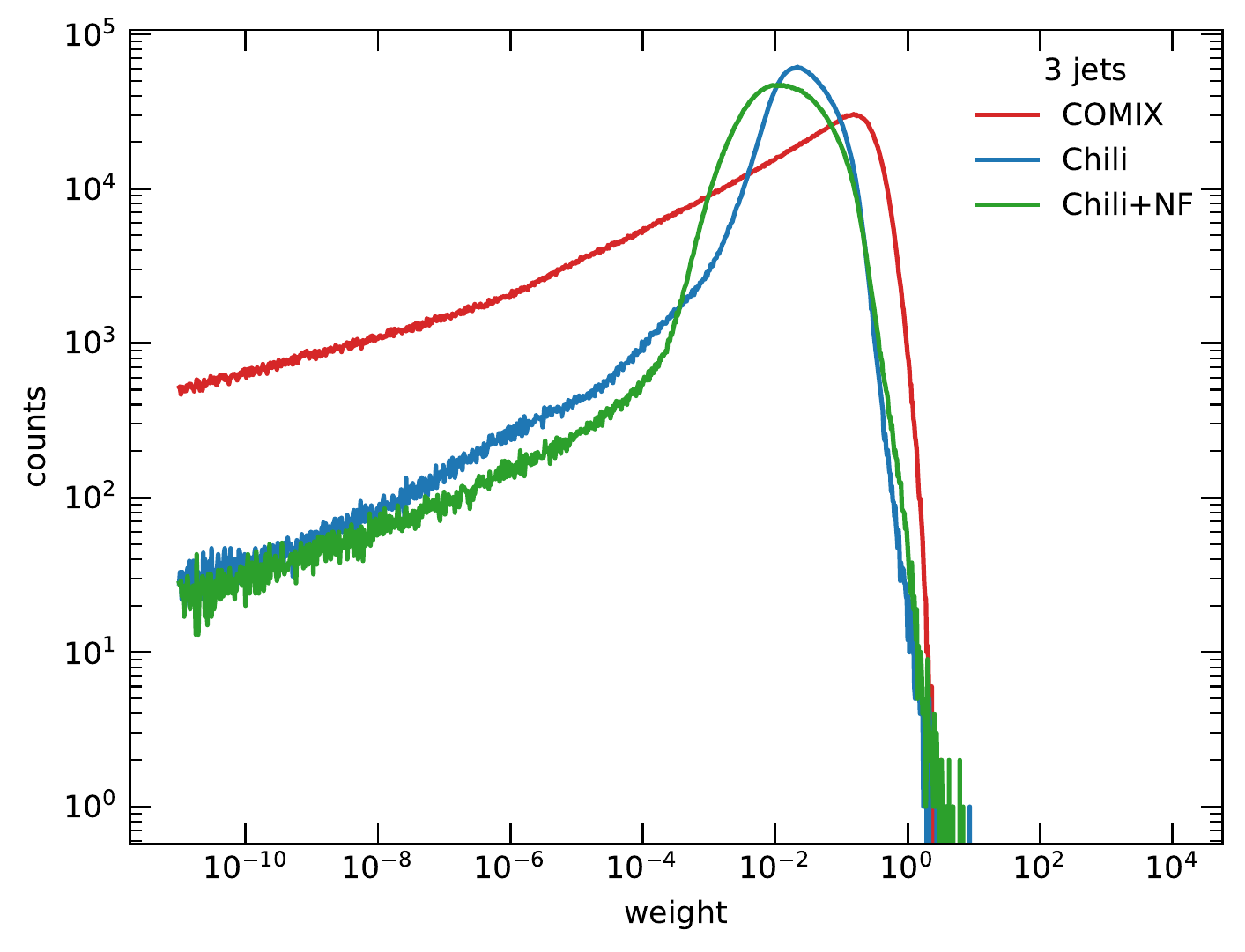}
    \caption{Same as Fig.~\ref{fig:comparison_madnis1}, but with an additional jet for each process.}
    \label{fig:comparison_madnis2}
\end{figure}
Figures~\ref{fig:comparison_madnis1} and~\ref{fig:comparison_madnis2}
show the weight distributions from 6 million phase-space points after
training for the simplest and next to simplest of our test processes.
We compare the recursive integrator of \Comix, \Chili with Vegas and 
\Chili in combination with \MadNIS. All results have been computed using
color summed matrix elements. It can be seen that the normalizing flow
based integrator yields a very narrow weight distribution in most cases.
A narrow weight distribution leads to good unweighting efficiencies as shown in
Tab.~\ref{tab:comparison_madnis}.
However, the default \Comix integrator
leads to a sharper upper edge of the weight distribution in the more
complex scenarios of Fig.~\ref{fig:comparison_madnis2}, which is more favorable
for unweighting. This indicates that the multi-channel approach with 
additional s-channels is favorable at high multiplicities.
We will investigate further this effect using the technology
developed in Ref.~\cite{Heimel:2022wyj}.
Furthermore, while the variance loss is optimal for achieving a narrow weight distribution,
by attempting to minimize the variance of the weight distribution.
However, this tends to result in a symmetric distribution about the mean.
This in turn leads to a less sharp upper edge
in the weight distribution, which results in a sub-optimal unweighting efficiency.
Additionally, the number of points required to reach optimal performance
for the normalizing flow is significantly higher than the Vegas based approaches, as demonstrated
in Ref.~\cite{Gao:2020vdv}. A study of the effect on the choice of loss function and other
hyper-parameters involved in the normalizing flow approach is left to a future work to
improve the unweighting efficiency at higher multiplicities and the convergence of the integrator.

\section{Outlook}
\label{sec:outlook}
We have presented a new phase-space generator that combines various existing techniques
for hadron collider phase-space integration into a simple and efficient algorithm.
This new integrator is not meant to become a replacement of the existing, tried and tested
techniques in state of the art parton-level event generators. Instead, we aimed at a simple,
yet practical solution with good Monte-Carlo efficiency, that offers the possibility to build
a scalable framework for event generation and can easily be ported to computing architectures
other than CPUs. Our new algorithm satisfies this requirement, because its
computational complexity scales linearly or at most polynomially with the number 
of external particles, and the complexity of the mapping can easily be adapted 
to the problem at hand. We have implemented the method in a scalable framework 
for CPU computing. Several extensions of this framework are in order: It should be
ported to allow the usage of GPUs. Computing platforms other than CPUs and GPUs could
be enabled with the help of Kokkos~\cite{kokkos} or similar computing models.
This becomes particularly relevant in light of recent advances in computing 
matrix elements on GPUs using portable programming models~\cite{
  Bothmann:2021nch,Valassi:2021ljk,Valassi:2022dkc,Bothmann:2022itv}.
In addition, the techniques for real-emission corrections should be extended beyond
\Sherpa, in order to make our generator applicable to a wider range of problems.
We also plan to further explore the combination of our new techniques with existing
neural-network based integration methods.

\section*{Acknowledgments}
We thank John Campbell for many stimulating discussions and his support of the project.
This research was supported by the Fermi National Accelerator Laboratory (Fermilab),
a U.S.\ Department of Energy, Office of Science, HEP User Facility.
Fermilab is managed by Fermi Research Alliance, LLC (FRA),
acting under Contract No. DE--AC02--07CH11359.
The work of F.H., S.H.\ and J.I.\ was supported by the U.S. Department of Energy,
Office of Science, Office of Advanced Scientific Computing Research,
Scientific Discovery through Advanced Computing (SciDAC) program, 
grant ``HPC framework for event generation at colliders''.
F.H. acknowledges support by the Alexander von Humboldt foundation.
E.B. and M.K.\ acknowledge support from BMBF (contract 05H21MGCAB). 
Their research is funded by the Deutsche Forschungsgemeinschaft 
(DFG, German Research Foundation) -- 456104544; 510810461.

\appendix
\section{Phase-Space efficiency}
\label{App:ps_efficiency}
Classical Monte Carlo unweighting relies on finding 
the maximum weight $w_{\mathrm{max}}$ during an inital optimization phase. Thereafter,
every Monte Carlo weight is compared against this maximum weight in a procedure called unweighting.
However, the procedure is prone to outliers in the weight distribution with the
potential to drastically reduce the unweighting efficiency and thus also the
compute efficiency. 
The procedure we used, as introduced in Ref.~\cite{Gao:2020zvv}, aims to reduce
the impact of outliers. In the following we briefly recall the algorithm:
\begin{enumerate}
    \item For $N_{\mathrm{Opt}}$ point in the last optimization step, 
    generate $n$ sets of events, each with $N_{\mathrm{Opt}}$ points.
    \item From these events, choose $m$ times $N_{\mathrm{Opt}}$ event samples and determine the maximum weight for each of them.
    \item Define $w_{\mathrm{max}}$ as the median of 
    the maximum weight of each of the sets.
\end{enumerate}
The numbers $n,m$ are to be chosen such that the unweighting efficiency stabilizes. In our case, we choose $n=m=100$.

\section{Recursive phase-space generator}
\label{sec:recursive_ps}
An efficient way for phase-space generation inspired by the diagram-based techniques
in~\cite{Byckling:1969sx} is given by the recursive phase-space generator introduced
in~\cite{Gleisberg:2008fv}. It relies on a matching of the basic building blocks for
the differential phase space to the Berends-Giele recursion.

Consider the $2\to n$ differential phase space in Eq.~\eqref{eq:n_particle_ps}.
According to Eq.~\eqref{eq:split_ps}, it can be factorized, where $\pi=\{1,\ldots,m\}$
corresponds to a set of particle indices. If we denote a subset of all possible particle
indices by greek letters, we can apply Eq.~\eqref{eq:split_ps} repeatedly to decompose
the complete phase space into basic building blocks corresponding to the $s$-channel 
production factor $(2\pi)^4\delta^4(p_\alpha+p_b-\sum_i p_i)$ and the two-body decays
${\rm d}\Phi_{2}(\alpha,b;\pi,\{a,b,1,\ldots,n\}\setminus\{\alpha,b,\pi\})$ and
${\rm d}\Phi_{2}(\pi;\rho,\pi\setminus\rho)$. These objects can be matched 
to the three-particle vertices occurring in the tree-level matrix element, 
as long as the particle index $b$ is held fixed.
Similarly, the integral ${\rm d}s_\pi/2\pi$, introduced in Eq.~\eqref{eq:split_ps}, 
can be matched to an s-channel propagator. It is then possible to show that the 
phase-space weight for a multi-channel integrator replicating the structures 
present in the Berends-Giele recursion can be computed using the same recursive algorithm. 

A key advantage of this recursive phase-space generator is that the computational
complexity scales at most exponentially with the number of outgoing particles,
while for diagram-based algorithms it scales factorially. More details on the
algorithm, including a simple example, can be found in~\cite{Gleisberg:2008fv}.

\bibliography{main}
\end{document}